\documentclass[%
 aps,
 prl,
twocolumn,
superscriptaddress,
frontmatterverbose, 
showpacs,preprintnumbers,
 nofootinbib,
 amsmath,
 amssymb,
floatfix,
latexsym,array,enumerate,letter,
]{revtex4-1}

\usepackage[utf8]{inputenc}
\usepackage[english]{babel}
\usepackage{multirow}
\allowdisplaybreaks
\usepackage{graphicx,color}
\usepackage[colorlinks=true,citecolor=blue,linkcolor=blue,urlcolor=blue]{hyperref}
\usepackage{comment}
\usepackage{appendix}
\usepackage{caption}
\usepackage{subcaption}
\usepackage{footmisc}
\usepackage{bm}
\usepackage{dcolumn}
\usepackage{gensymb}
 \usepackage{float} 
\restylefloat{table}
\usepackage{multirow}
 \usepackage{hyperref}
 \usepackage{array}

\hypersetup{
    colorlinks,
    citecolor=black,
    filecolor=black,
    linkcolor=blue,
    urlcolor=blue
}

\newcommand{\eps}{\epsilon}

\newcommand{\beq}{\begin{equation}}
\newcommand{\eeq}{\end{equation}}
\newcommand{\bcn}{\begin{center}}
\newcommand{\ecn}{\end{center}}
\newcommand{\omg}{\omega}

\newcommand{\Omg}{\Omega}
\newcommand{\ga}{\gamma}
\newcommand{\al}{\alpha}
\newcommand{\te}{\theta}

\begin{document}


\title{Rotational Properties of Inverted Hybrid Stars}

\author{Rodrigo Negreiros}
 \email{rnegreiros@id.uff.br}
\affiliation{%
 Catholic Institute of Technology \\ 1 Broadway - 14th floor, Cambridge, MA 02142
}%
\affiliation{%
  Instituto de Física, Universidade Federal Fluminense \\ - Av. Litoranea - Niteroi - RJ - BR 
}%


\author{Chen Zhang}
\affiliation{
 The HKUST Jockey Club Institute for Advanced Study, The Hong Kong University of Science
and Technology, Hong Kong SAR, People’s Republic of China\\
}%

\author{Renxin Xu}
\affiliation{
 Department of Astronomy, School of Physics, Peking University, Beijing 100871,  People’s Republic of China\\
}%


\date{\today}

\begin{abstract}
We study the rotational properties of inverted hybrid stars (IHSs), which have been recently proposed as a possible new class of compact stars characterized by an outer layer of quark matter and a core of hadronic matter, in an inverted structure compared to traditional hybrid stars. 
We analyze distinct models representing varying depths of quark-hadron phase transitions. Our findings reveal that, while IHSs rotating at their Kepler frequencies typically exhibit a significantly higher mass and larger circumferential radius as anticipated, interestingly, there is a significant increase in potential twin configurations in the case of rapid rotations.  
We further study sequences of constant baryonic mass, representing potential paths of rotational evolution. Our results indicate that not all stars in these sequences are viable due to the onset of phase transitions during spin-down, leading to possible mini-collapses. We also investigate the phenomenon of ``back-bending" during spin-down sequences, which is manifested in a rather distinct shape for IHSs due to their inverted structure and the large density discontinuity caused by the strong phase transition.
Our study extends to the braking index of IHSs, which could serve as observational hints for the detection of such objects. Furthermore, we also consider sequences of IHSs with constant angular momentum and demonstrate that they exhibit, with relatively high confidence, an approximate universal relation between supramassive configurations and the spherical mass limit. Furthermore, we also show that $\hat{I}-\hat{Q}$ relations found in previous studies also hold approximately for IHSs to some extent. Our research enriches existing studies by introducing the significant aspect of rotation, unveiling intriguing phenomena that could serve as observational markers.

\end{abstract}

\maketitle


\section{\label{sec:Intro}Introduction}

The concept of phase transitions within compact stars is a very realistic possibility that has been seriously considered for decades \cite{doi:10.1142/S0218301318300084,orsaria2019phase,heiselberg1999phase,glendenning2001phase,yang2008influence,reddy2000first,glendenning1992first,burgio2002hadron}. Typically, based on our understanding of the baryonic matter phase diagram, the transition from hadronic matter (HM) to deconfined quark matter (QM) at some point within the compact object is considered, with the precise point of this transition is contingent upon the equation of state (EOS) of both the hadronic and quark phases \cite{alford2005hybrid,zhang2023hybrid,cierniak2021hybrid,peng2008deconfinement,blaschke2010hybrid,dexheimer2012hybrid,dexheimer2013hybrid}. A significant challenge in the study of hybrid stars is determining the hadronic-quark matter transition at the low temperatures typical of compact stars. Furthermore, it remains to be conclusively determined whether the transition occurs at the densities found within the interiors of neutron stars, although it has been suggested that this is probable \cite{annala2020evidence}. Compact stars that undergo a phase transition to quark matter at inner regions are commonly referred to as hybrid stars.

 Recently, the provocative possibility of ``inverted" hybrid stars was proposed and examined~\cite{Zhang2023,Zhang2024}. These objects are composed of an outer layer of quark matter, with the inner stellar region composed of ordinary hadronic matter, in opposite to what one expects for traditional hybrid stars, hence the name ``inverted" hybrid stars (IHSs) or cross stars by other names. This concept is supported by the hypothesis of absolutely stable QM at low pressure, as demonstrated by \cite{Witten:1984rs} for strange quark matter (SQM) that is composed of a nearly equal number of $u,d,s$ quarks, and \cite{Holdom2018} for up-down quark matter ($ud$QM) that composed of $u,d$ quarks\footnote{Following~\cite{Holdom2018}, the properties of $ud$QM are in active explorations~\cite{Iida:2020fyt,Xia:2020byy,Xia:2022tvx,Bai:2024muo,Bai:2024amm,Brown:2024gqu}.}, together with the possibility that at high pressures hadronic matter may be more stable than QM due to a smaller chemical potential in a large physical EOS parameter space~\cite{Zhang2023}. This inverted phase transition cannot be excluded since it resides in the intermediate density and low-temperature regime where the QCD dynamics is non-perturbative and lattice calculation suffers from the fermion sign problem.  Besides, as \cite{Zhang2023} observed, composition with $ud$QM allows a larger parameter space than those with SQM.
 As SQM can form strange quark stars~\cite{1986ApJ...310..261A, Weber:2004kj,Zhang2021, Pretel:2024pem,Zhou:2024syq}, $ud$QM can also form up-down quark stars~\cite{Zhang:2019mqb,Ren:2020tll,Cao:2020zxi, Wang:2021byk,Li:2022vof,Yuan:2022dxb}. IHSs can be seen as configurations in which a hadronic core developed inside these quark stars.

The properties of non-rotating (spherically symmetric) IHSs have been thoroughly examined in~\cite{Zhang2023}, with a focus on their macroscopic features such as mass and radius. This study also explored key quantities for potential merger scenarios, including tidal deformability. Later, the asteroseismology of IHSs was systematically studied in~\cite{Zhang2024}. Building on the findings of \cite{Zhang2023}, our current research delves deeper into the attributes of these objects, with a particular emphasis on the properties of rotating IHSs. The characteristics of rotating compact stars have garnered significant interest recently, particularly due to the potential emergence of high-angular momentum compact stars from merger events. We argue that an in-depth study of rotating IHSs is justified, as it could yield valuable insights for future dynamic simulations of post-merger remnants, in addition to its inherent scholarly value.

This paper is organized as follows: Section I introduces the context and purpose of this work; Section II provides an overview of the microscopic model used in this study; Section III discusses our findings on the structure of rotating IHSs; Section IV explores potential evolutionary sequences of rotating stars; Section V describes the back-bending phenomena as observed in rotating cross stars; and Section VI offers our conclusions and future perspectives. In this paper, we adopt units where $G=c=1$.

\section{Microscopic Model}

In this study, we adopt the methodology of \cite{Zhang2023} and explore the concept of IHSs, which are theorized to be composed of quark matter at lower densities and hadronic matter at higher densities. As discussed in the introduction, we focus on the IHSs with $ud$QM, which has the grand potential expressed as follows:
\begin{equation}
    \Omega = -\frac{\xi_4 a_4}{4 \pi^2} \mu^4 + B, \label{gPOT}
\end{equation}
where $\mu$ is the average quark chemical potential.
This can be understood as the grand potential of a free quark gas with QCD corrections parameterized by the parameter $a_4$, with $a_4 = 1$ signifying no corrections and smaller values of $a_4$ denoting larger corrections \cite{Alford2005}. $B$ is the effective bag constant. The parameter $\xi_4$ for $ud$QM, is given by \cite{Zhang2023} $\xi_4 = \left[ (1/3)^{4/3} + (2/3)^{4/3} \right]^{-3} \approx 1.86
$.

From the potential given in Eq.~(\ref{gPOT}) we can write the following equations for the energy per baryon of $ud$QM \cite{Zhang2021}
\begin{equation}
  (E/A)_q = \frac{3\sqrt{2 \pi}}{(a_4 \xi_4)^{1/4}}B^{1/4},
\end{equation}
and the chemical potential
\begin{equation}
\mu_{\rm Q}=\frac{3\sqrt{2\pi}}{(a_4 \xi_4)^{1/4}}(P+B)^{1/4},
\end{equation}
while the equation of state is given by
\begin{equation}
    P = \frac{\epsilon - 4B}{3},
\end{equation}
with $\epsilon$ denoting the energy density. Note that while the EOS for $ud$QM is independent of $a_4$, it will affect the chemical potential $\mu_Q$ and thus the point of quark-hadron phase transition. 

The $ud$QM parameters must be defined within the $ud$QM hypothesis. As discussed in \cite{Zhang2023} we must take into account the absolute stability condition of QM at zero pressure, which constrains $(B, a_4)$ via $(E/A)_q<930$ MeV, the energy per baryon number for ${}^{56}$Fe, known to be the nucleus with the lowest mass per nucleon. Therefore, following in the footsteps of \cite{Zhang2023}, we consider the following benchmark values for the bag constant $B = 20, 35, 50$ MeV/fm$^3$ with the corresponding values of $a_4 = 0.35, 0.62, 0.88$, respectively. Each pair of values for parameters $B$ and $a_4$ will be denoted, heretofore as parameters set A, B, C. Explicitly, set A corresponds to $B = 20$ MeV/fm$^3$ and $a_4 = 0.35$, set B to $B = 35$ MeV/fm$^3$ and $a_4 = 0.62$, and set C to $B = 50$ MeV/fm$^3$ and $a_4 = 0.88$.

\begin{figure}[h]
    \centering
    \includegraphics[width=1.0\linewidth]{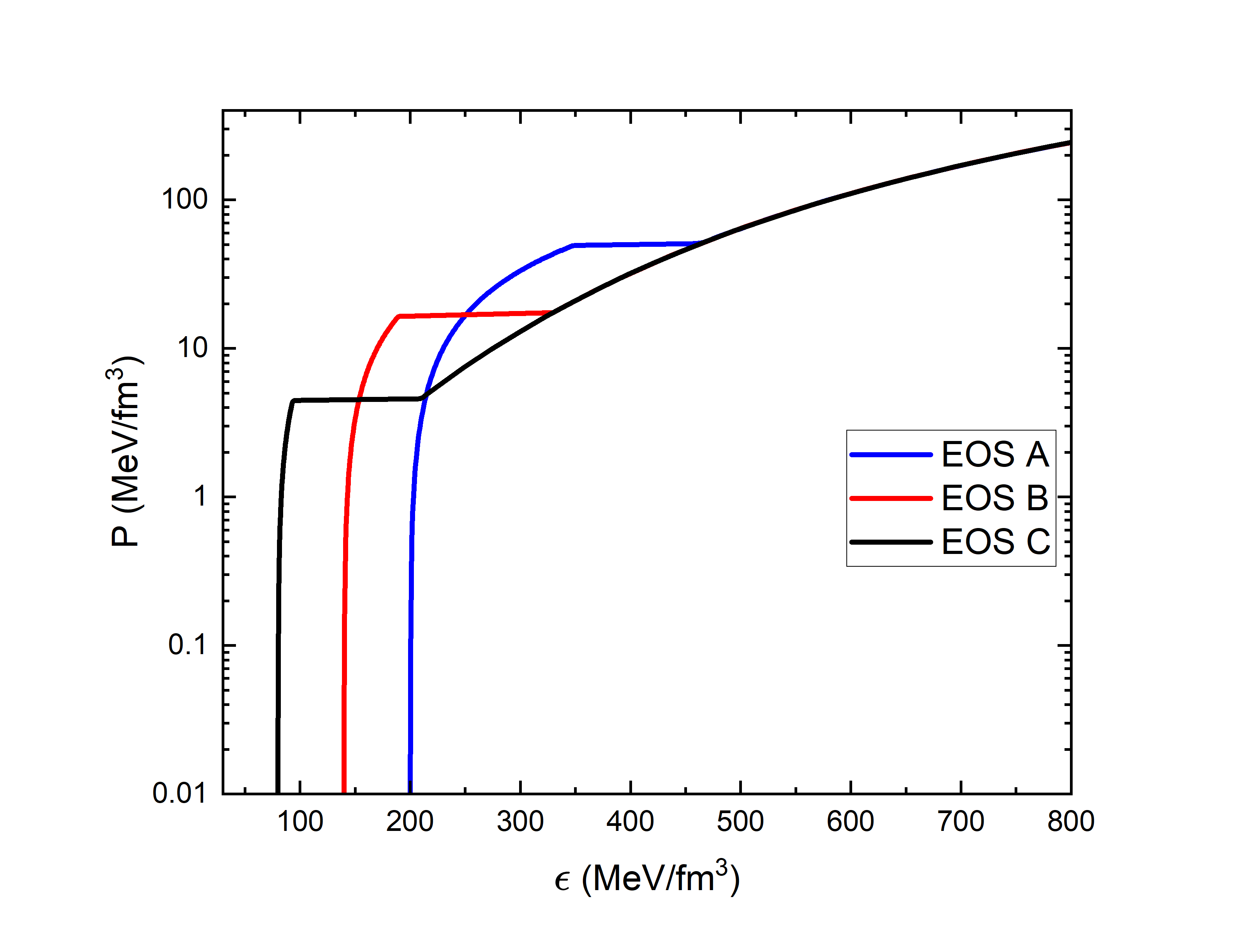}
    \caption{The EOS for inverted hybrid stars involves $ud$QM transitioning to hadronic matter (APR model) at high pressure. Parameter sets A, B, and C represent varying values for the $ud$QM bag constant and the $a_4$ parameter. Specifically, set A corresponds to $B = 20$ MeV/fm$^3$ and $a_4 = 0.35$, set B to $B = 35$ MeV/fm$^3$ and $a_4 = 0.62$, and set C to $B = 50$ MeV/fm$^3$ and $a_4 = 0.88$.}
    \label{fig:EOS}
\end{figure}

In the study of IHSs, it is necessary to consider a hadronic EOS. The authors of \cite{Zhang2023,Zhang2024} performed a comprehensive examination of a representative set of hadronic equations of states, which revealed similar qualitative features and small quantitative differences (for details, see their appendices). Consequently, following their observations, in this work we consider the APR EOS \cite{Akmal1998} for the hadronic sector, as sourced from the CompOSE database \cite{COMPOSE}. Employing the $ud$QM hypothesis, as well as the aforementioned hadronic EOS, quark matter transits to hadronic matter as pressure increases beyond the transition pressure $P_{\rm cr}$ as determined by the crossing of the chemical potentials of the two matter phases. The combined EOS for each benchmark model of IHSs is shown in Fig.~\ref{fig:EOS}.

\section{Structure of Rotating stars}

The structure of rotating neutron stars presents a significantly more complex problem to solve compared to their non-rotating counterparts. For non-rotating, spherically symmetric neutron stars, the combination of Einstein’s equation and the equilibrium condition yields the renowned Tolman-Oppenheimer-Volkoff Equations (TOV) \cite{Tolman1939,Oppenheimer1939}.

However, the complexity escalates when rotation is introduced into the equation. This is primarily due to the absence of an external, analytical solution for the exterior of a rotating neutron star. This contrasts with the case of spherically symmetric stars, where Birkhoff’s theorem demonstrates that the external solution aligns with Schwarzschild’s solution for a point mass.

In the case of a rotating neutron star, it becomes necessary to solve Einstein’s equation across all space, applying the appropriate boundary conditions at infinity. For a rotating star exhibiting axial symmetry around its rotation axis, we select a metric that maintains the requisite symmetry. The metric is given by
\begin{equation}
\begin{split}
 ds^2 =& g^{\mu \nu}dx_\mu dx_\nu  =\\ &= -e^{\ga + \rho} dt^2 + e^{2\al}(dr^2 + r^2d\te^2)\\ & + e^{\ga -\rho} r^2 \sin^2\te (d\phi - \omg dt)^2, \label{met_rns}
 \end{split}
\end{equation}
where the metric functions $\rho, \ga, \al$ and $\omg$ are function of $r$ and $\te$ only. The metric function $\omg$ represents the frame dragging frequency, i.e. the frequency at which the local inertial frames are rotating. 

The sources of curvature are specified in the energy-momentum tensor. We treat neutron star matter as a perfect fluid in which case the energy-momentum tensor is given by
\begin{equation}
 T^{\mu \nu} = (\eps + P)u^\mu u^\nu + P g^{\mu \nu}, \label{EMT}
\end{equation}
where $\eps$ is the energy density, $P$ is the pressure, and $u^\mu$ is the four-velocity. 

The field equations of the metric functions are obtained by substituting the metric (\ref{met_rns}) and the energy-momentum tensor (\ref{EMT}) into Einstein's field equation
\begin{equation}
G^{\mu \nu} = R^{\mu \nu} -\frac{1}{2}g^{\mu \nu} =  8\pi T^{\mu \nu}, \label{Einstein_FE}
\end{equation}
where $R^{\mu \nu}$ is the Riemman curvature tensor.

To find the stellar structure one must solve Eq.~(\ref{Einstein_FE}) in addition to the hydrostatic equilibrium condition, which can be obtained from imposing conservation of momentum-energy ($T^{\mu \nu}~_{;\mu} = 0$), yielding
\begin{equation}
 dP - (\eps + P)[d\ln u^t + u^t u_\phi d\Omg] = 0, \label{HYD_rns}
\end{equation}
where $\Omg$ is the matter's angular velocity. For rigid rotations, as in this paper, one has $d\Omg = 0$. 

These equations must then be solved in all space, with flat space at infinity as boundary conditions. In this work we employ the code \textit{Astreus}, which has been widely used to study rotating neutron stars \cite{2017A&A...603A..44N,Negreiros14,2013PhLB..718.1176N,2012PhRvD..85j4019N,negreiros2010m}. This code employs the numerical method described in \cite{Komatsu1987} and have been thoroughly explored in \cite{Stergioulas1995b}. 

\begin{figure}[h!]
    \centering
    \includegraphics[width=1.0\linewidth]{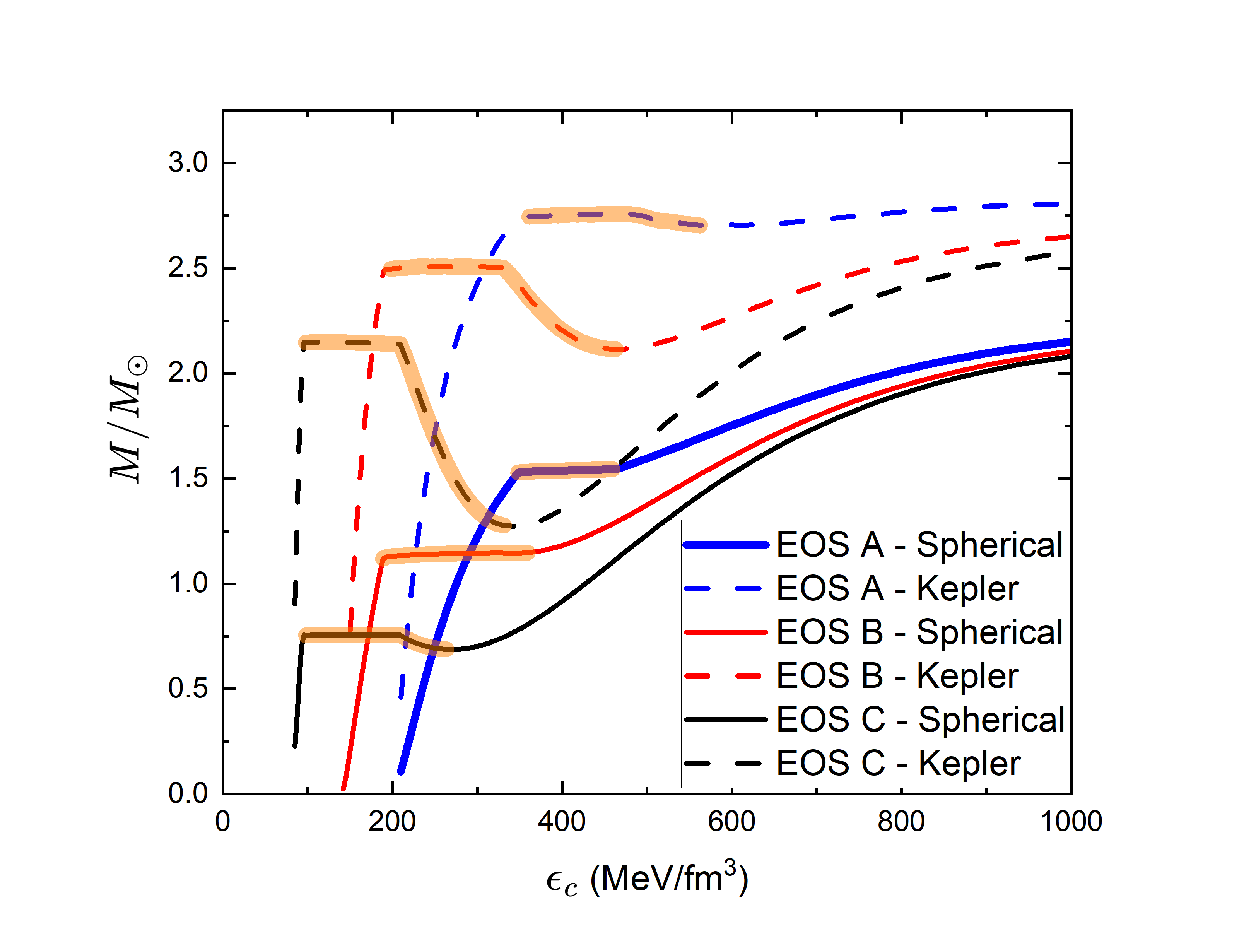}
    \caption{Mass as a function of central density is depicted for three different Equations of State (EOS) of compact stars (IHSs) studied. For each EOS, two curves are presented, represented by solid and dashed lines, corresponding to the spherical sequence where stars have no rotation, and the Kepler sequence where each star rotates at its maximum (Kepler) frequency, respectively. Orange shaded regions represent unstable configurations}
    \label{Fig:Mxec}
\end{figure}

We show in Fig.~\ref{Fig:Mxec} the mass as a function of the central density corresponding to the three different EOSs for IHSs that we have considered.

For each EOS examined, we present two curves: one representing the spherical sequence with non-rotating stars (TOV solution), and the other depicting the Kepler sequence where each star rotates at its maximum (Kepler) frequency, that is, the highest frequency before the onset of mass shedding in stars.

The findings illustrated in Fig.~\ref{Fig:Mxec} reveal that IHSs undergoing a transition from quark matter to hadronic matter at lower densities tend to exhibit smaller masses. Conversely, transitions occurring at higher densities result in larger masses. 

As anticipated, the impact of rapid rotation is significant, with stars that rotate at the Kepler frequency exhibiting considerably higher mass than their non-rotating counterparts.

It is also worth noting the emergence of twin star configurations, especially for IHSs where the transition occurs at lower densities. Twin star configurations refer to scenarios where two stable stars from the same family have identical mass but differ in central densities (and radii), resulting in varying compactness \cite{schramm2016modelling,dexheimer2015role,lyra2023compactness}. Typically, twin star setups involve the star with the smaller central density (less compact) being purely hadronic, while its more compact counterpart is hybrid. However, the case of inverted hybrid stars has the reversed situation, where the less compact twin is composed solely of quark matter, and the more compact twin contains both quark matter and hadronic matter, as explicitly shown in~\cite{Zhang2023} for non-rotating IHSs. Here our new finding reveals that, when rotation is introduced, an intriguing phenomenon emerges: the twin branch significantly expands. This is particularly noteworthy in the case of EOS B, where a relatively pronounced twin branch appears when rotation is factored in.
This phenomenon may also appear in traditional hybrid stars. For instance, Ref.~\cite{Tsaloukidis:2022rus} showed that adding rotations can introduce twin stars from normal stars.

\begin{figure}[h!]
\centering
    \includegraphics[width=\linewidth]{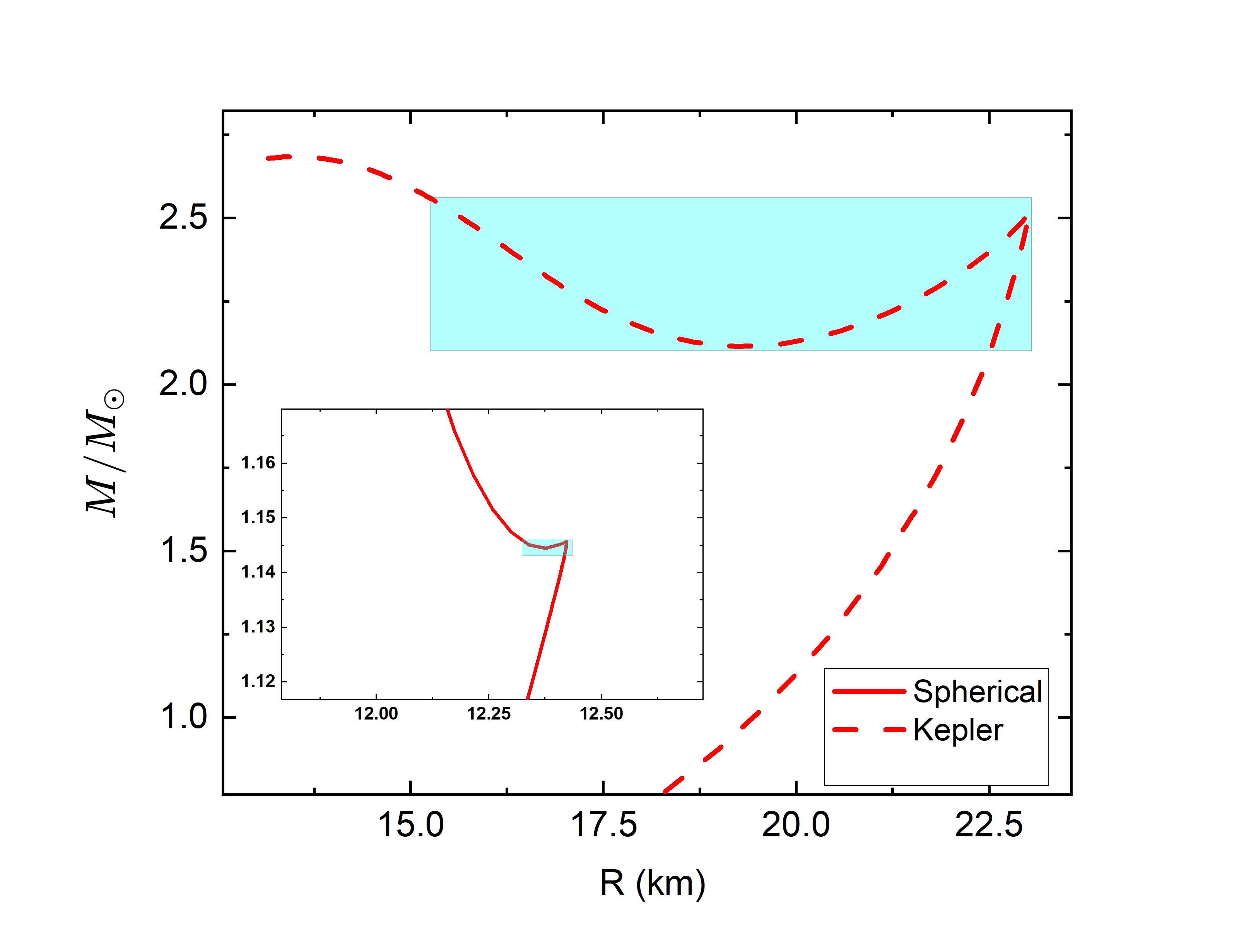}
       \includegraphics[width=\linewidth]{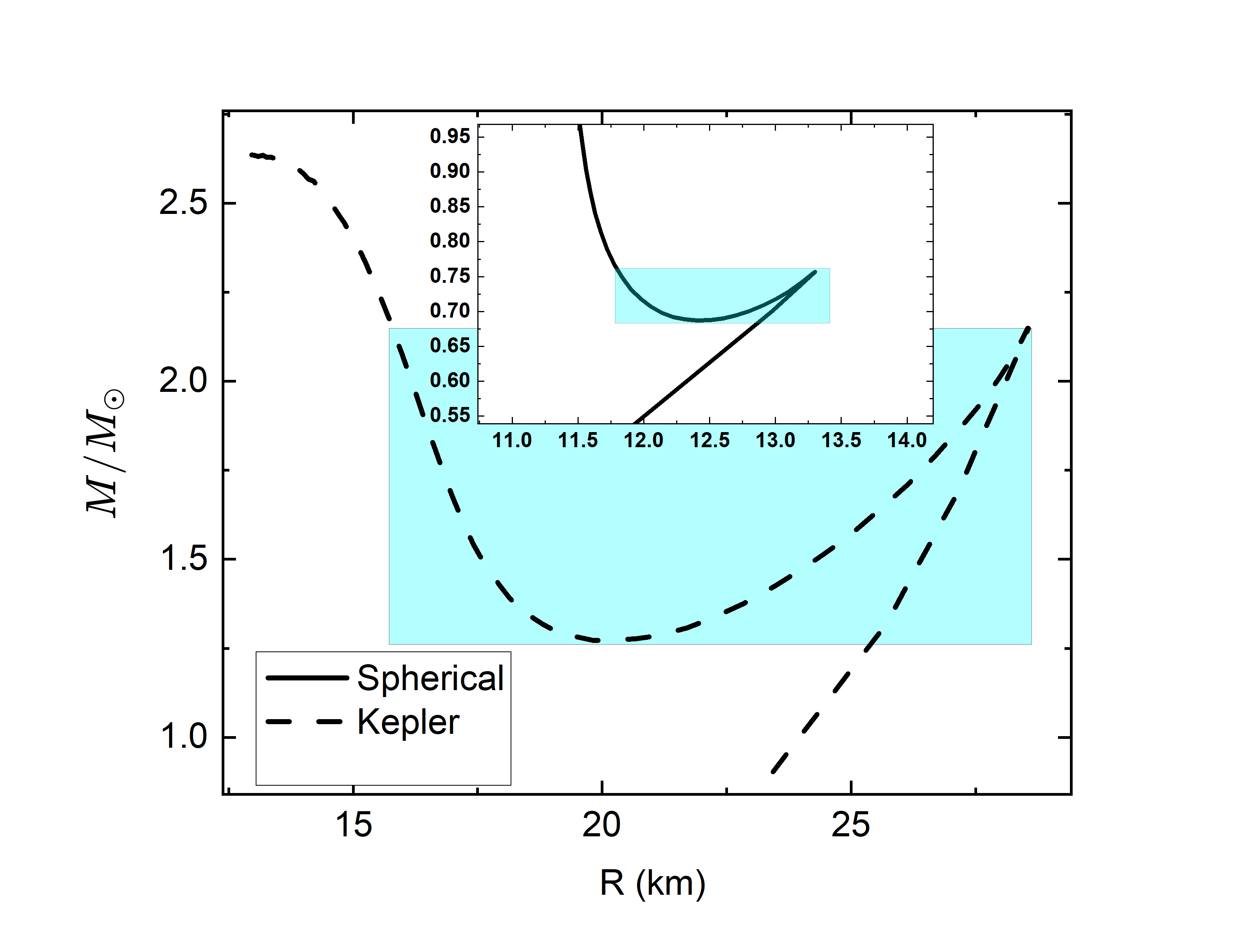}
    \caption{The mass as a function of circumferential radius for IHSs of EOS B (top panel) and EOS C (bottom panel) rotating at their Kepler frequencies. The blue shaded area denotes the zone where twin configurations occur. The insets illustrate the twin configurations of corresponding spherically symmetric (non-rotating) stars. Orange-shaded regions represent unstable configurations.}
    \label{fig:MxR_EOSB}
 \end{figure}

The emergence of twin-star configurations becomes more pronounced when examining the Mass-Radius diagram for these IHSs sequences. These findings are depicted in Figures~\ref{fig:MxR_EOSB}, which display the gravitational mass as a function of the circumferential radius. The cyan-shaded region denotes the twin-star configurations. The insets in these figures illustrate the twin configurations for the corresponding non-rotating ones of the same EOS.
 
The findings outlined in Figures~\ref{fig:MxR_EOSB} highlight a marked increase in twin configurations. The non-rotating scenario reveals minimal twin configurations, while the Kepler frequency sequence exhibits a broad spectrum of twin stars. Even though the Kepler sequence represents an extreme case, there are also intermediate scenarios, underscoring that the inclusion of angular momentum can promote the emergence of twin configurations.

Previous studies have also explored the possibility of twin stars in rotating configurations. \cite{Largani2022} investigated various models for rotating stars, including the novel aspect of considering finite temperatures. They observed the emergence of twin configurations; however, in their study, this phenomenon occurred with hot matter, rather than as a result of rotation, which is the focus of our present study. Another significant investigation into twin configurations in rotating stars is \cite{Bozzola2019}. This work considered several microscopic models for hybrid stars, including those with a first-order phase transition. Notably, the models they explored exhibited prominent twin-star configurations in both rotating and nonrotating sequences. Despite their exemplary and comprehensive study of rotating hybrid stars, their findings do not indicate the emergence of twin stars solely due to rotation.

\section{Evolution Sequences}
Given that a compact star may be born with a specific angular momentum, either after a supernova or a merger event, exploring potential rotational evolution paths as they decelerate is warranted. This involves examining the evolution of a sequence with constant baryon mass from a state of high rotation to the nonrotating limit. Such sequences can reveal the potential evolutionary trajectories of isolated stars that decelerate without mass loss or gain, thereby preserving a constant baryonic mass.

\begin{figure}[H]
    \centering
    \includegraphics[width=1\linewidth]{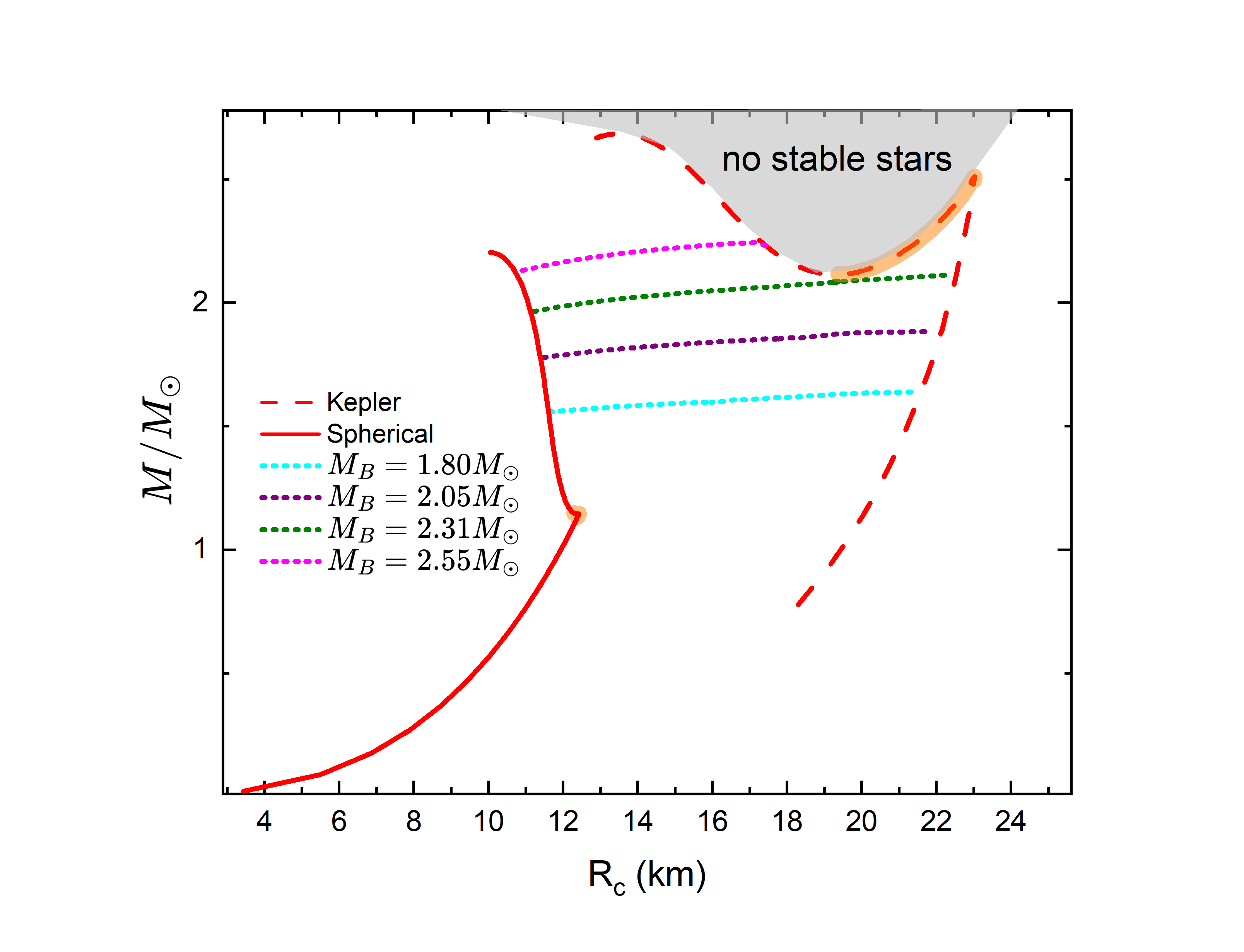}
    \caption{Gravitational mass as a function of circumferential radius for IHSs of EOS B. Each dashed curve represents evolution paths for stars with the indicated baryon mass. The gray-shaded region depicts the forbidden region above the Kepler sequence. Similarly, orange shaded region represents unstable configurations. }
    \label{fig:MxRseq_B}
\end{figure}

To illustrate such sequences, we show in Fig.~\ref{fig:MxRseq_B} the mass as a function of circumferential radius for three sequences of stars with constant baryonic mass, namely $M_B = 1.80,~ 2.05,~ 2.31$ and $2.55\, M_\odot$, as well as the spherical and Keplerian families for EOS B, with the gray-shaded region depicting the forbidden region above the Keplerian sequence. These sequences illustrate possible paths taken by stars with constant baryonic mass, i.e. without losing or gaining matter, as they spin down from a high angular momentum configuration (up to the limit set by the Kepler sequence) to a non-rotating configuration represented by the spherical sequence. The findings in the various EOSs examined are observed to be qualitatively consistent. This implies that all the studied EOSs display the same characteristics. To prevent an overload of figures, we have chosen to illustrate only the results for EOS B, hence, only the results for EOS B are illustrated to avoid an excessive number of figures. 
It is crucial to note that while the stars depicted by the constant baryonic sequences in Fig.~\ref{fig:MxRseq_B} are hydrostatically stable, not all may be stable against oscillations. The stability condition for a sequence of stars with constant baryonic mass follows the turning point theorem (see for instance \cite{haensel2016rotating}) and can be written as
\begin{equation}
    \left. \left( \frac{\partial J}{\partial \epsilon_c} \right) \right\rvert_{M_b = ~ \textrm{const.}} < 0,
    \label{eq:jstab}
\end{equation}
where $J$ is the angular momentum and $\epsilon_c$ the central energy density. We note that, strictly speaking, the condition given in eq.~(\ref{eq:jstab}) should be used to determine secular stability, which may not necessarily coincide with dynamical stability \cite{Friedman1988}. Besides the aforementioned condition, it is also necessary to consider the energy density gap in the EOS, which is evident in Fig.~\ref{fig:EOS}, resulting from the phase transition. 

\begin{figure}[H]
    \centering
    \includegraphics[width=1\linewidth]{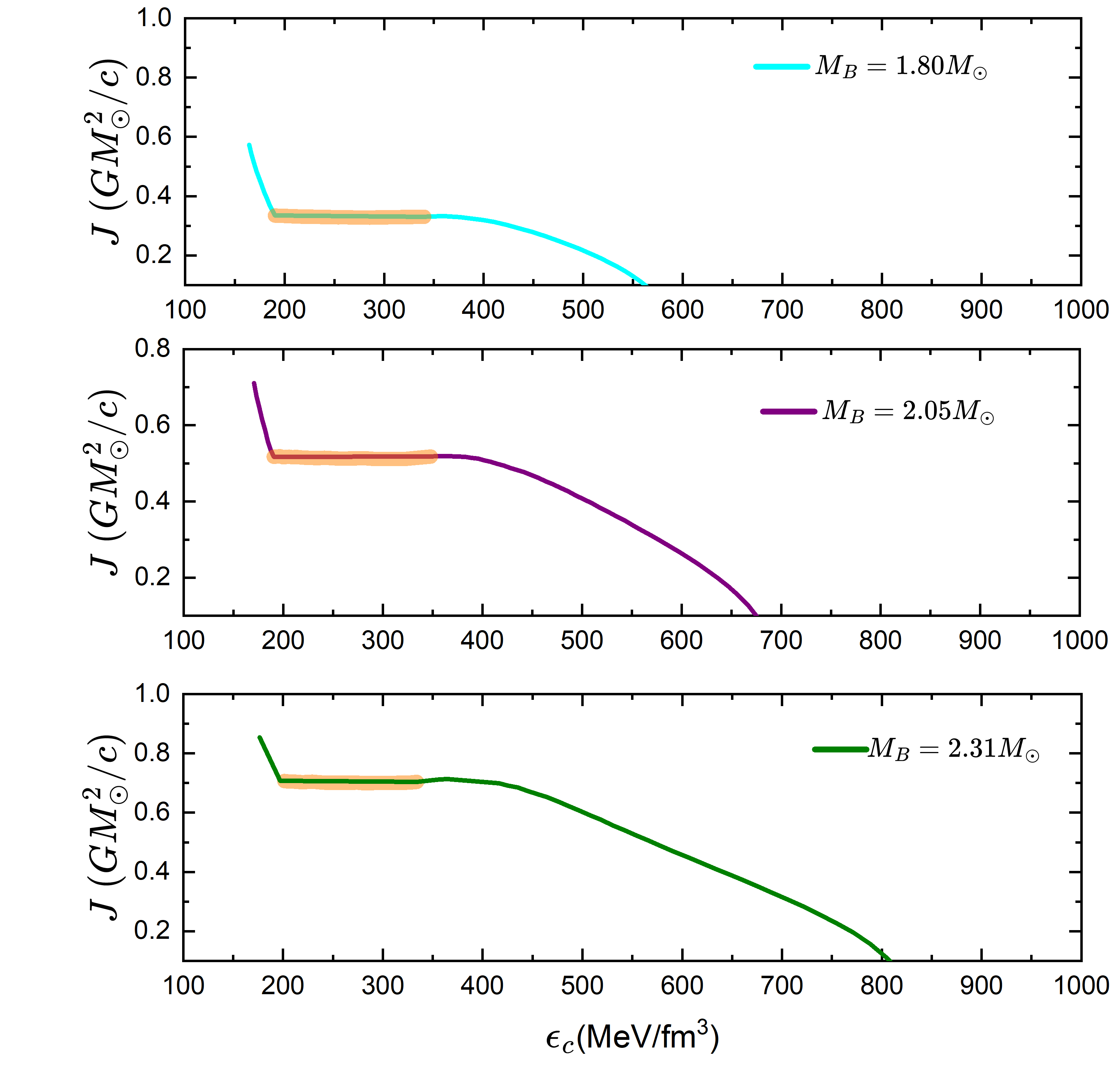}
    \caption{Angular momentum as a function of central density for the constant baryonic mass sequences indicated. The orange highlighted region represents forbidden either due to the phase transition gap and/or stability criteria for constant baryon mass sequences.  } \label{fig:Jxec}
\end{figure}

Taking into account these two constraints, we illustrate in Fig.~\ref{fig:Jxec} the relationship between angular momentum and central density for the three constant baryonic mass sequences depicted in Fig.~\ref{fig:MxR_EOSB}. In these figures, we highlight the forbidden regions, as defined by the condition (\ref{eq:jstab}), and the prohibited region resulting from the phase-transition gap.

\begin{figure}[H]
    \centering
    \includegraphics[width=1\linewidth]{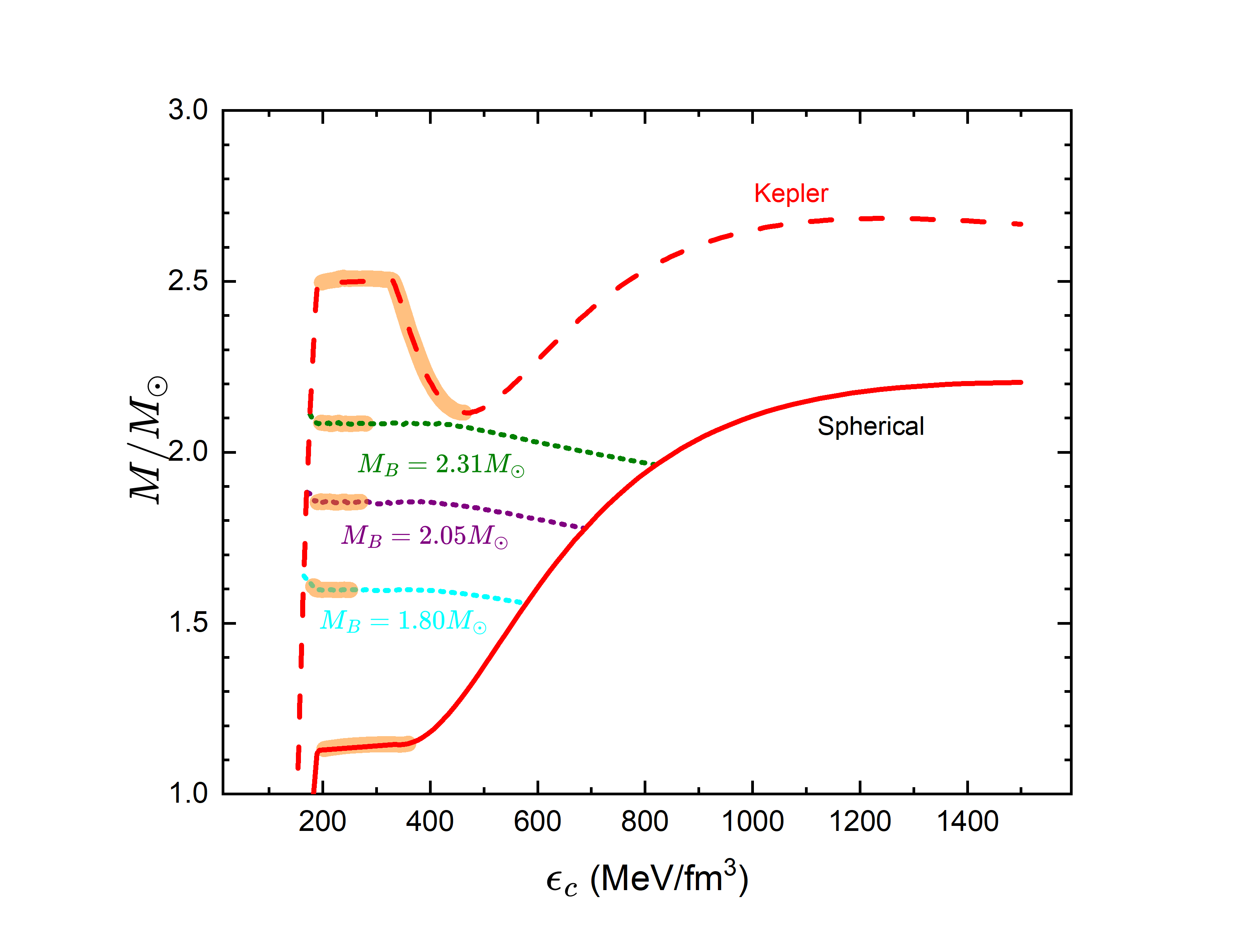}
    \caption{Gravitational mass as a function of central density for the sequence of stars of EOS B. Black and red solid lines represent spherical and Kepler sequences, respectively. Dashed curves represent constant baryon mass sequences with the indicated baryonic mass, and orange-shaded regions represent forbidden configurations. } \label{Fig:Mxec_B}
\end{figure}

We can now achieve a better understanding of the potential evolutionary paths of stars with constant baryonic mass by analyzing such sequences in the mass versus central density diagram depicted in Fig.~\ref{Fig:Mxec_B}.

\begin{figure}[H]
    \centering
    \includegraphics[width=1\linewidth]{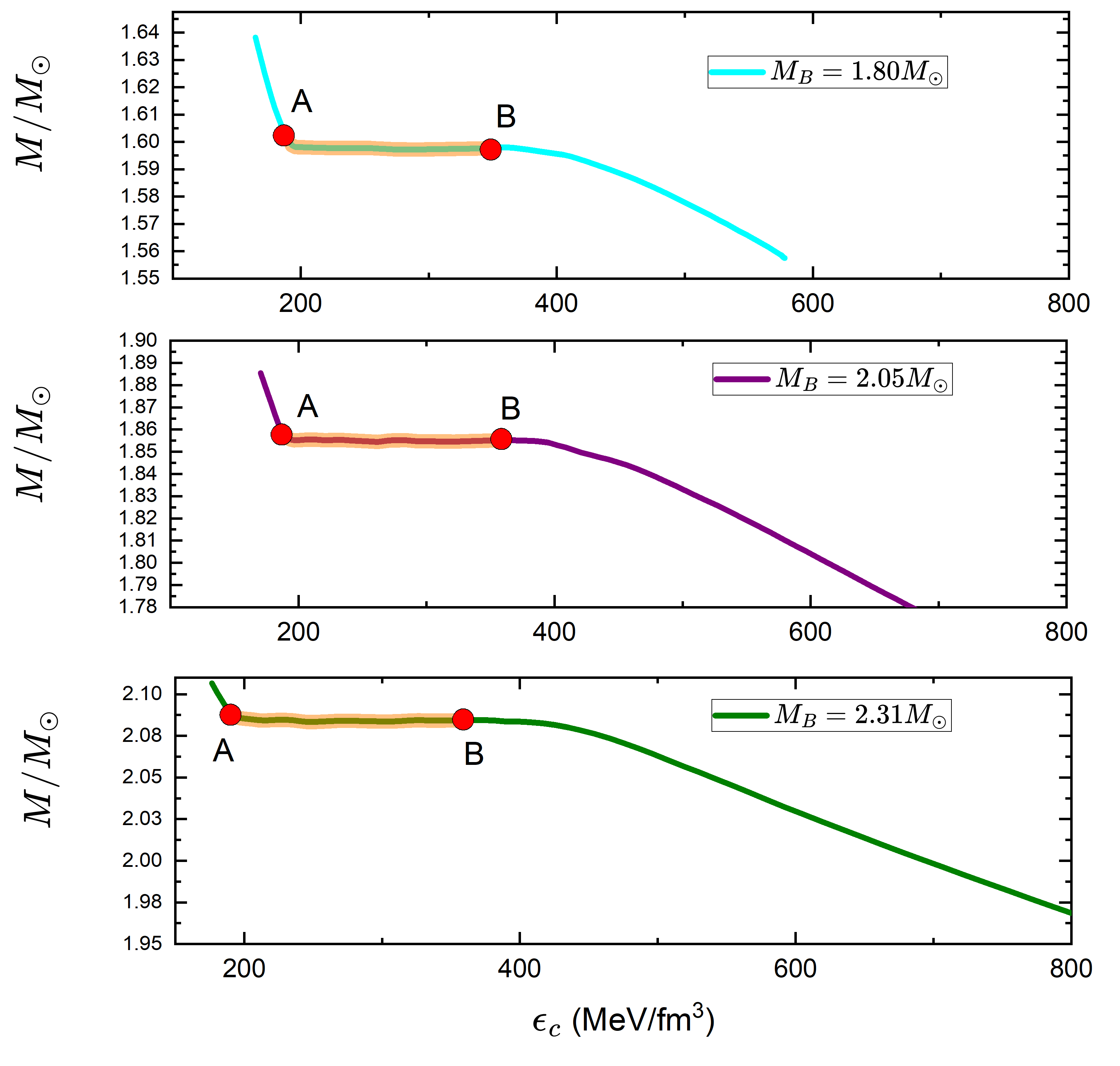}
    \caption{Zoomed-in baryonic mass sequences shown in Fig.~\ref{Fig:Mxec_B}. Points A and B in each figure denote the last stable configuration before and after the forbidden region indicated by the orange shaded region.} \label{Fig:Mxec_B_zoom}
\end{figure}

As observed in Fig.~\ref{Fig:Mxec_B}, there exists a small region of unstable configuration near  the Kepler sequence. This suggests that stars may originate at lower frequencies (beyond the orange-shaded regions). Alternatively, if they are born with frequencies close to the Kepler limit, they will experience a sudden transition through the prohibited region. A more detailed analysis of the constant baryonic mass curves can provide a better understanding of this phenomenon, which can be facilitated with the aid of Fig.~\ref{Fig:Mxec_B_zoom}

Fig.~\ref{Fig:Mxec_B_zoom} exhibits a closer look at the constant baryonic mass sequences of Fig.~\ref{Fig:Mxec_B}. It also indicates two points, $A$ and $B$, which represent the two stable configurations at the boundary of the forbidden regions. Any star whose initial configuration is before point $A$, will undergo a fast transition from point $A$ to $B$. This must be an exothermic process, as the star will lose gravitational mass (mostly due to angular momentum loss) in such a process as $M_B < M_A$. This difference of energy must be partly released as electromagnetic and/or gravitational radiation, whereas the rest will act as heating. 

Considering that baryonic mass is constant for the sequences under consideration, we can easily estimate the amount of energy released when a star goes from point A to point B as
\begin{equation}
    E_g = M_g^A - M_g^B,
\end{equation}
where $E_g$ is the energy released, $M_g^A$ and $M_g^B$ are the gravitational mass at points A and B, respectively. Taking the sequence of $M_B = 1.80\, M_\odot$ for instance, we obtain
\begin{equation}
    E_g \approx 2.4\times10^{-3} M_\odot \approx 4.3 \times 10^{51}\, \textrm{erg}.
\end{equation}

A similar result was outlined in reference \cite{haensel2016rotating}. Similarly, references \cite{Chubarian2000, Mellinger2017} explored the possibility of phase transitions triggered by spin-down. However, in those cases, high-frequency stars in a purely hadronic phase underwent a phase transition to quark matter during spin-down. In contrast, our study finds that high-frequency stars are solely composed of quarks, with a phase transition to hadronic matter occurring during spin-down. The work of \cite{Ayvazyan2013} also concluded that heat would be released during spin-down triggered evolution; however, in their study, the phase transition was to a color-superconducting quark phase.

One must bear in mind, however, that not all of this energy is to be radiated away, as part of the energy will remain in the star in the form of heat. Even if only 1\% of the energy is radiated it could give rise to very energetic phenomena such as gamma-ray bursts and fast-radio bursts~\cite{Wang:2024opz}, thus warranting further investigations on the dynamics of such phenomena.


\section{Back-Bending}

\begin{figure}[H]
    \centering
    \includegraphics[width=1\linewidth]{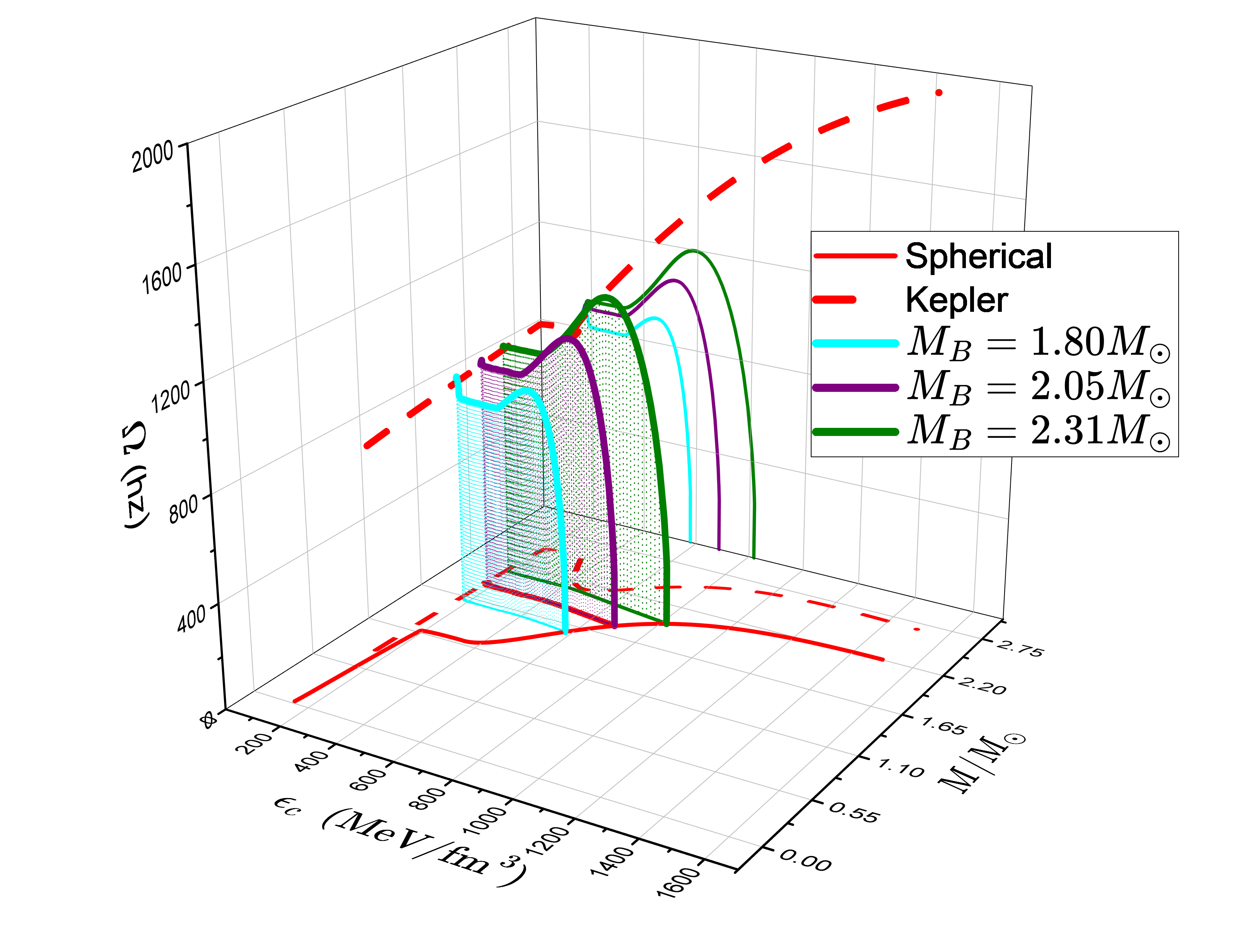}
    \caption{3D representation of the stellar sequences shown in Fig.~\ref{Fig:Mxec_B}, with the $z$-axis representing the rotational frequency. Solid black and red lines represent spherical and Kepler sequences, respectively. Cyan, purple and green curves represent the constant baryonic mass sequences. Also shown in the graph are the projection on the three perpendicular planes, as well as droplines connecting the 3D curves to the $x-y$ plane.  } \label{fig:3dfreq}
\end{figure}

A typical manifestation of the possibility of a phase transition in the core of neutron stars is the phenomena known as ``back-bending" \cite{haensel2016rotating}. This phenomenon can be comprehended by considering the star’s contractions during its rotational evolution, which reduces the moment of inertia, in conjunction with the softening of the EOS that accompanies the phase transition. These combined factors lead to an increase in the star’s rotational frequency, despite a reduction in its total angular momentum during spin-down. This behavior can be distinctly identified by introducing an additional axis, representing the rotational frequency, to Fig.~\ref{Fig:Mxec_B}. The resulting graphic, depicted in Fig.~\ref{fig:3dfreq}, clearly demonstrates an increase in rotational frequency corresponding to an increase in stellar density during the star’s rotational evolution.

\begin{figure}[H]
    \centering
    \includegraphics[width=1\linewidth]{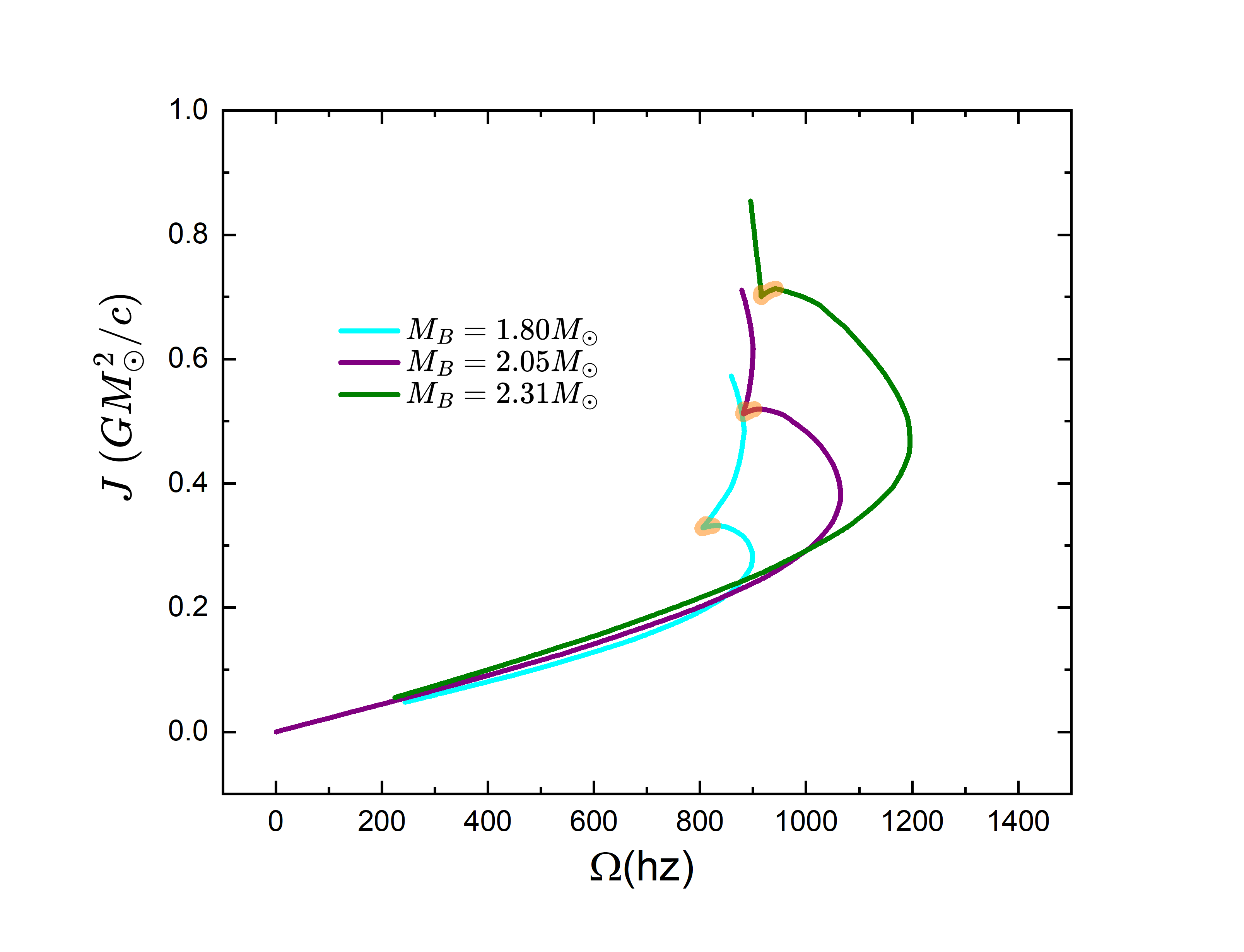}
    \caption{Angular momentum as a function of frequency for constant baryonic mass sequences of EOS B. Each colored curve follow constant baryonic mass as indicated in the figure. Orange shaded region represents unstable configurations.} \label{fig:Jxfreq}
\end{figure}

The increase in frequency illustrated in Fig.~\ref{fig:3dfreq} is associated with the so-called back-bending, which can be observed when the angular momentum is plotted against the frequency, as can be seen in Fig.~\ref{fig:Jxfreq}

The results depicted in Fig.~\ref{fig:Jxfreq} exhibit intriguing behavior. Although they retain the expected properties associated with the previously studied back-bending phenomenon (an increase in rotational frequency corresponding to a decrease in angular momentum), the shape of the $J$-$\Omega$ curve is markedly different. Traditional back-bending curves typically show an `S' shape, hence the name ‘back-bending’. However, for the inverted hybrid stars studied here, we observe a much sharper bend in the curve, more akin to a ‘back-kink’. We attribute this behavior to the nature of the inverted hybrid stars under study and the abrupt density change in nature when low-pressure $ud$ QM transitions to hadronic matter at higher pressures. 
We note that previous studies have also identified a similar behavior  as the  `back-kink,' as observed in \cite{zdunik2006phase, bhattacharyya2005rotating}. In these works, ordinary hybrid stars also exhibit a similar `back-kink,' which can be attributed to the significant discontinuity in density at the phase transition. A key advantage of the inverted hybrid star model is that, even with a sharp first-order phase transition and large density discontinuities, its \(M_{\rm TOV}\) is more likely to reach the desired \(2 M_{\odot}\) target (see Table I of Ref.~\cite{Zhang2024}). This allows for the existence of a more massive hybrid object (\(M > 2 M_{\odot}\)) that exhibits a more pronounced back-kink behavior than normal hybrid stars.
Although we currently have limited observational data regarding the angular momentum of compact stars, future observations of this nature could potentially provide empirical evidence for the model studied here.

\section{Braking Index}

A relevant observable quantity for pulsars is the braking index, which is defined in terms of the observable frequency ($\Omega$) and its first and second derivatives ($\dot{\Omega},\ddot{\Omega}$) and is written as \cite{Manchester1977,Lyne2012,Blandford1988,Melatos1997}
\begin{equation}
    n = \frac{\Omega\ddot{\Omega}}{\dot{\Omega}^2}.
\end{equation}

In the slow rotation approximation, the spin-down of a neutron star is normally assumed to take the form of a power law. In the ideal case where the only braking is due to the magnetic field, one obtains $n=3$. On the other hand, if the braking is due to gravitational wave emission one gets $n=5$ \cite{Lyne2012}. 

A more accurate description for the braking index of a relativistic, rapidly rotating neutron star is found in \cite{Spyrou2002} which can be written, for the case in which the spin-down is solely due to the magnetic field as
\begin{equation}
    n(\Omega) = 3 - \frac{2I'\Omega + I''\Omega^2}{I + I'\Omega}, \label{n_full}
\end{equation}
where a prime indicates a derivative with respect to the frequency ($\Omega$).

\begin{figure}[H]
    \centering
    \includegraphics[width=1\linewidth]{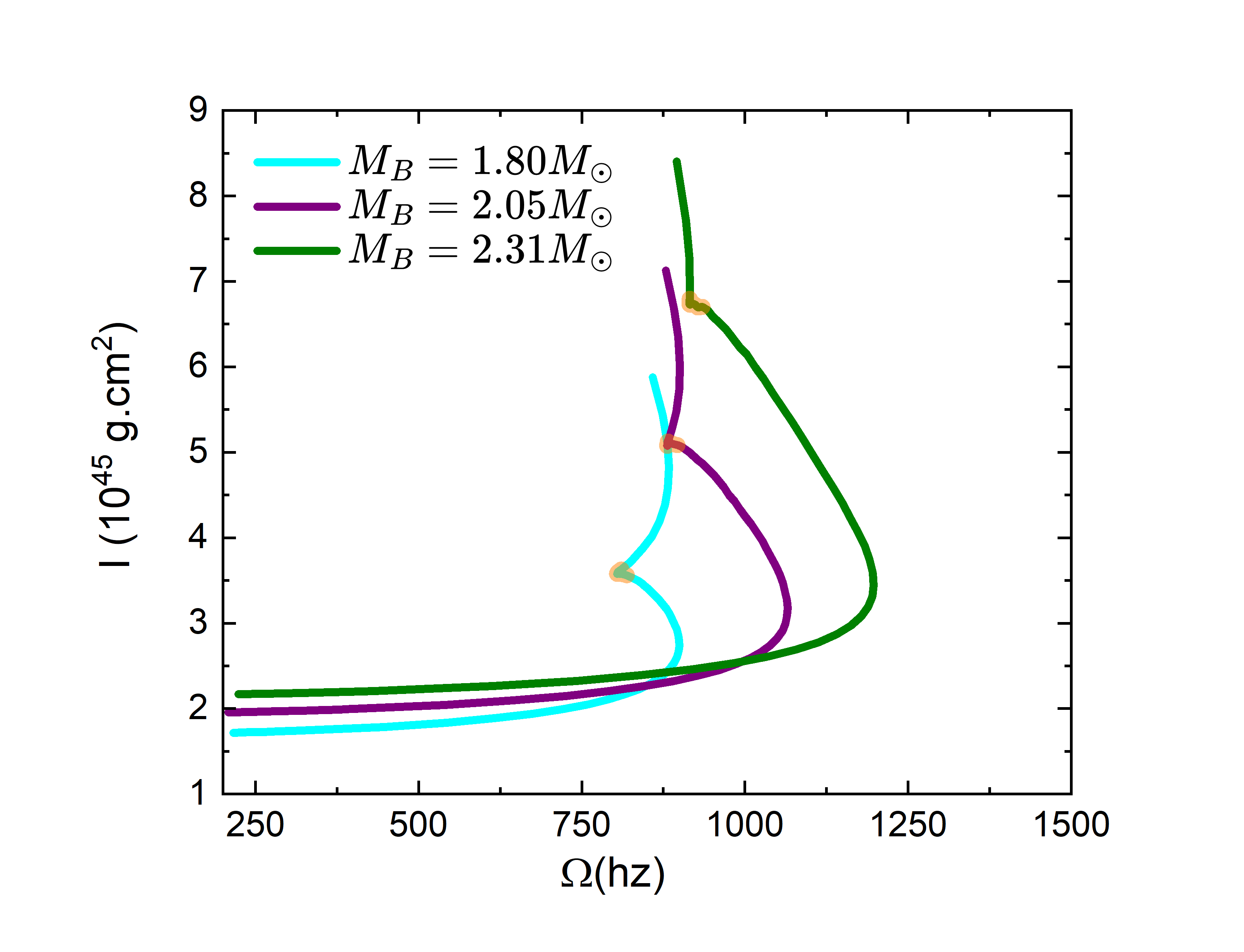}
    \caption{Moment of inertia as a function of rotation frequency for three representative stars of EOS B. Each curve represents the indicated baryonic mass sequence, as indicated on the figure. The orange-shaded region represents unstable configurations.} \label{Fig:Mom_x_Freq}
\end{figure}

The braking index can be considered an interesting way of probing the interior of pulsars, as observations can be compared with predicted values to provide hints about the stellar interior. We begin our analysis by studying the moment of inertia of the three representative stars studied in the previous section ($M_B = 1.80, 2.05$ and $2.31$ M$_\odot$). The dependence of the moment of inertia as a function of frequency is shown in Fig. (\ref{Fig:Mom_x_Freq}).

We note that all three stars exhibit similar qualitative behavior, with the moment of inertia smoothly growing for the stars with lower frequencies (higher densities) which have both hadronic and quark phases. The moment of inertia thus turns back at a maximum frequency, only to turn back again, in an effect mimicking the back-bending studied in the previous section (which in reality is a result of this behavior for the moment of inertia in stars with phase transition). 

\begin{figure}[H]
    \centering
    \includegraphics[width=1\linewidth]{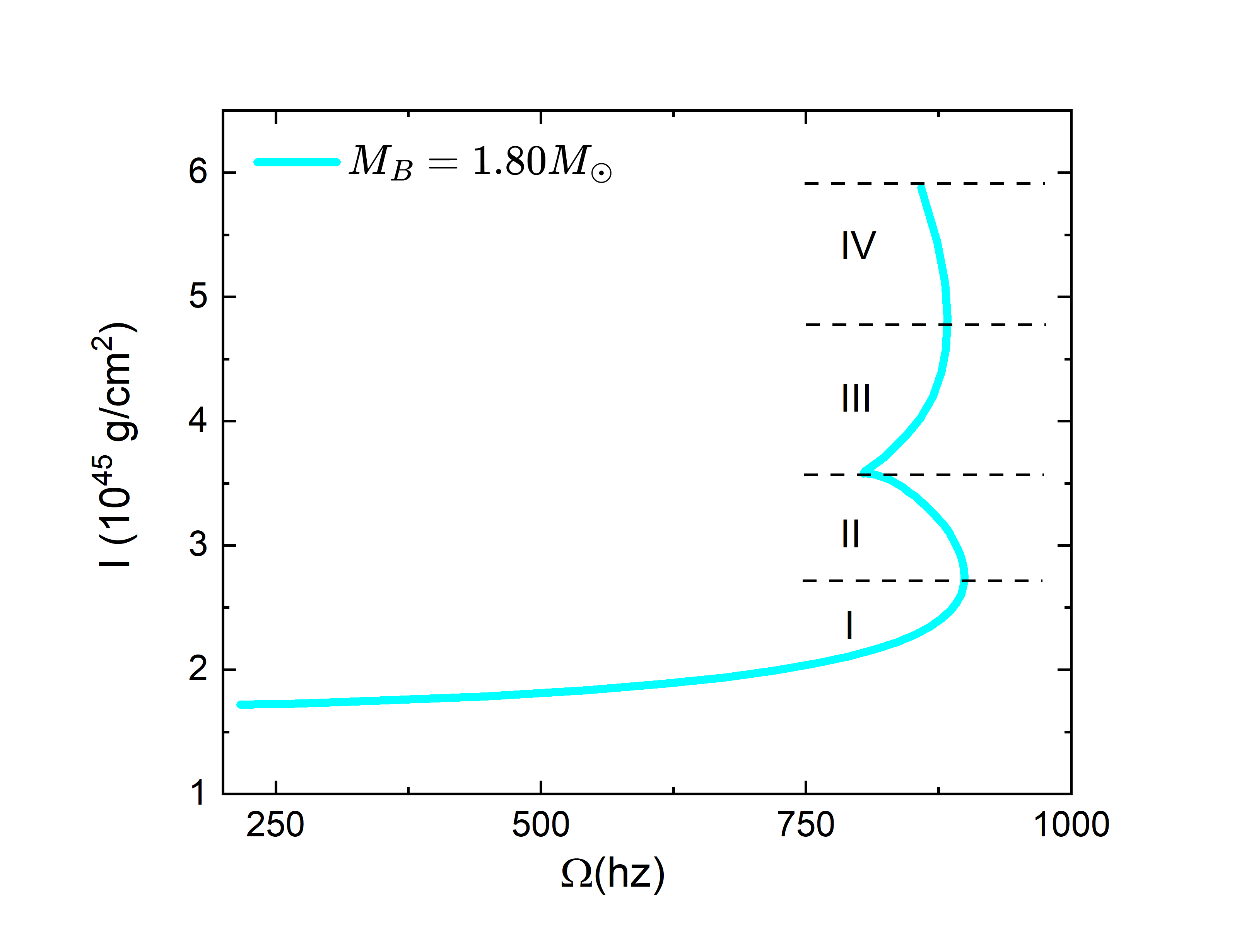}
    \caption{Moment of inertia as a function of rotation frequency the star with $M_B = 1.80$ M$_\odot$ of EOS B. Also shown are four different regions which will be used to study the star's braking index. } \label{Fig:M_x_freq_M180}
\end{figure}

Such behavior indicates that we can expect wildly different values for the braking index of stars with this equation of state, much like those stars studied in \cite{Spyrou2002}. To further investigate the braking index of these stars we now focus only on the $M_B = 1.80$ M$_\odot$, as the same qualitative behavior is found for the other stars. We show in Fig. (\ref{Fig:M_x_freq_M180}) the star's moment of inertia, with 4 separate regions that will be analyzed.

\begin{figure*}[t]
    \centering
    \includegraphics[width=1\linewidth]{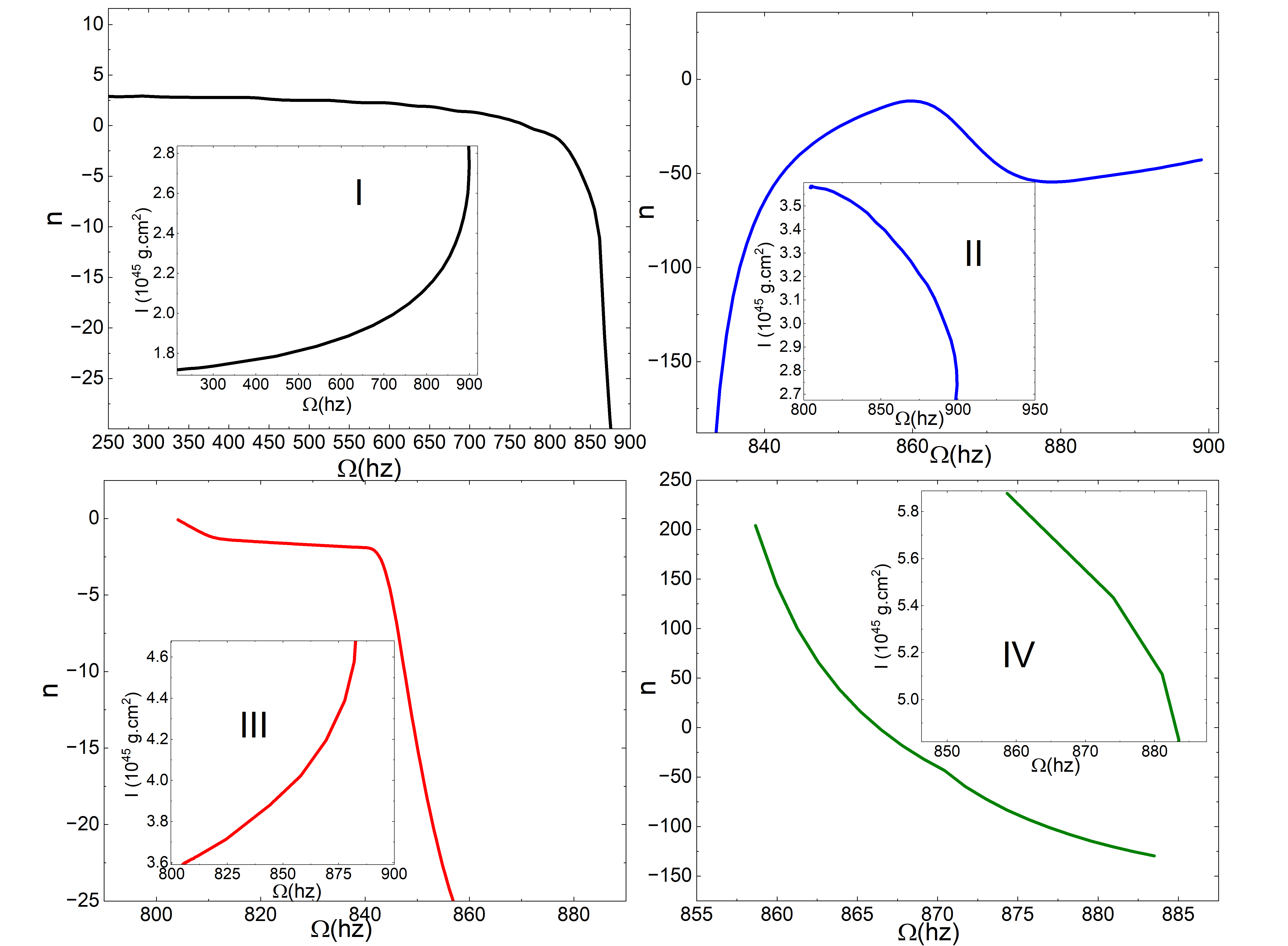}
    \caption{Braking index for the $M_B = 1.80$ M$_\odot$ divided in the the four regions indicated in  Fig.\ref{Fig:M_x_freq_M180}. The insets indicate the corresponding moment of inertia section.} \label{Fig:n_full}
\end{figure*}

Using the moment of inertia as a function of rotational frequency, shown in Fig. (\ref{Fig:M_x_freq_M180}), we use eq.~(\ref{n_full}) to calculate the braking index for the star in the four indicated regions. This result is shown in Fig. \ref{Fig:n_full}

Fig.~\ref{Fig:n_full} indicates that for lower frequencies, thus higher densities and lower deformations, the star has a breaking index near the canonically expected value of 3. As the frequencies increase and deformations start to take place, the non-linear aspects of the moment of inertia evolution become noticeable and we have the braking index taking large negative or positive values. These results are somewhat typical of stars with first-order phase transition in their interior, as discussed in \cite{Spyrou2002}, and may be useful for comparison with observed pulsars that may exhibit unusual values for the braking index, as this may be an indication of a phase transition to exotic phases.

\section{Universal Relations}

There has been great interest in studying EOS insensitive relations for rotating neutron stars, the so-called universal relations. Many papers have devoted great efforts to finding such relations \cite{Breu2016,Bozzola2019,Khadkikar2021}. We now discuss a few of these relations, particularly those associated with supramassive configurations, as well as the relationship between the moment of inertia and quadrupole moment ($I-Q$), which holds particular relevance for the study of merger events.

\begin{figure}[h]
    \centering
    \includegraphics[width=1\linewidth]{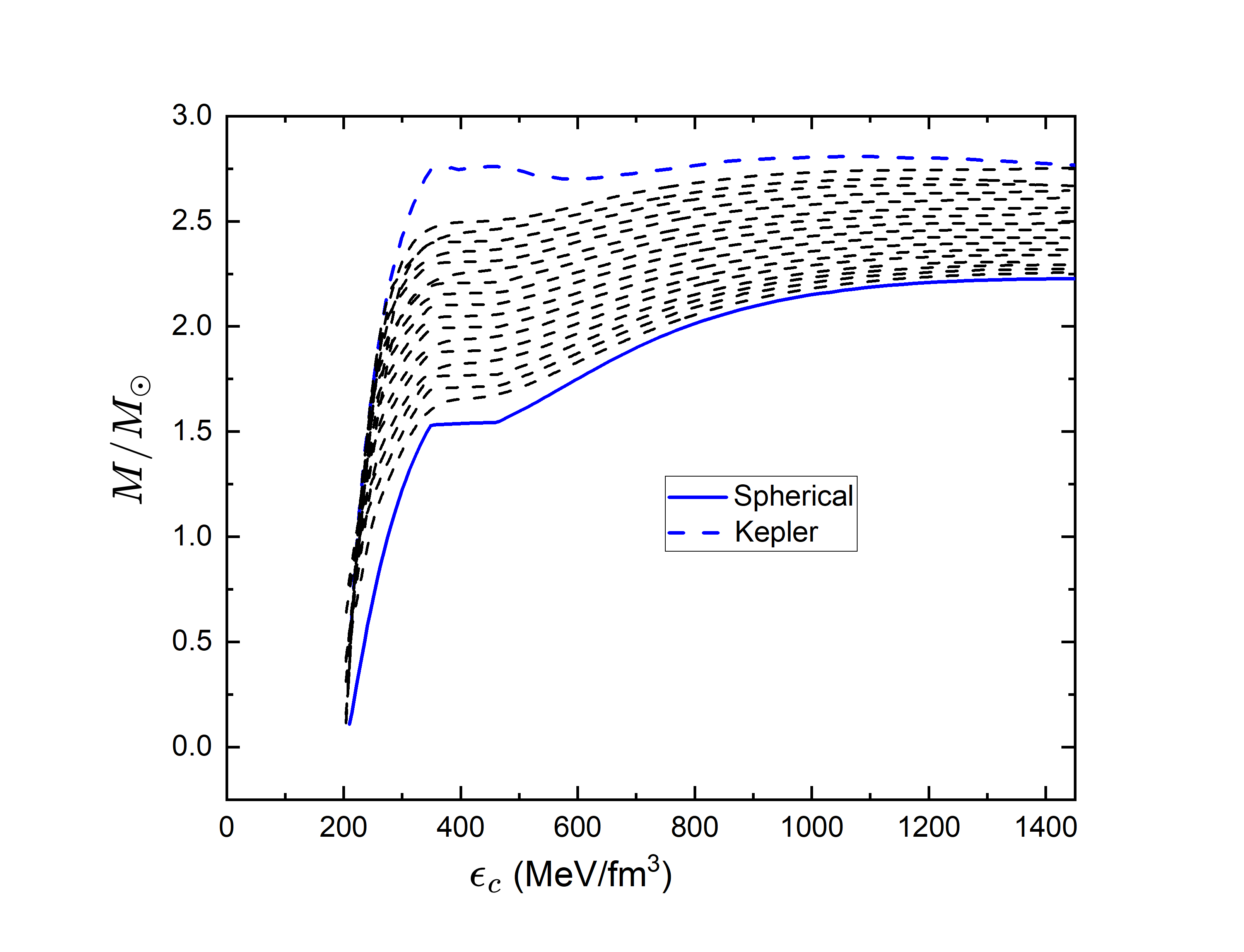}
    \caption{Sequence of stars with constant angular momentum for EOS A. The solid black line represents spherically symmetric stars with zero angular momentum, whereas the red curve represents the Kepler sequence, in which stars rotate at their maximum possible frequency. Dashed lines represent sequences with constant angular for a range of values between zero and maximum allowed rotation. } \label{fig:JseqEOSA}
\end{figure}

\begin{figure}[h]
    \centering
    \includegraphics[width=1\linewidth]{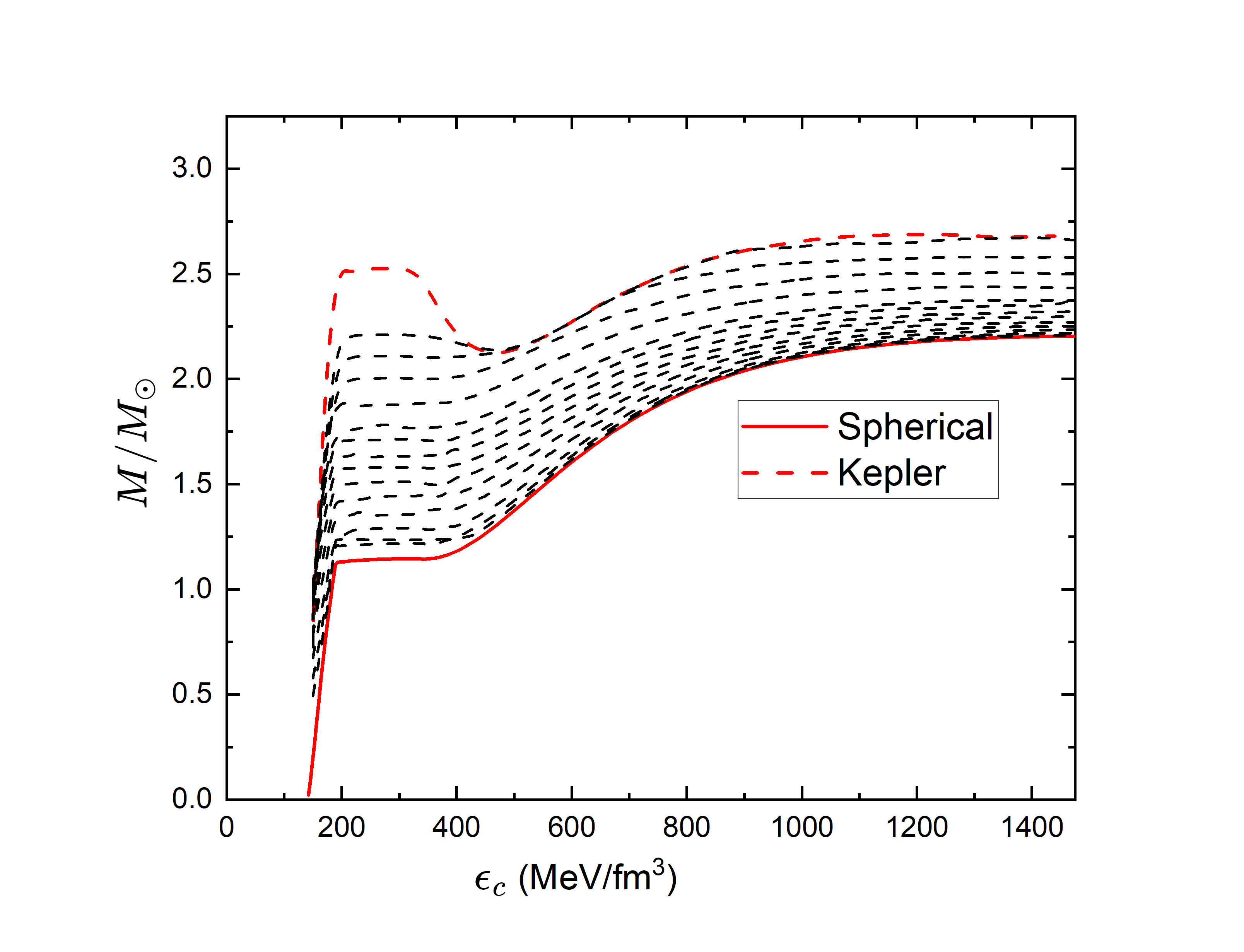}
    \caption{Same as Fig.~\ref{fig:JseqEOSA} except for EOS B. } \label{fig:JseqEOSB}
\end{figure}

\begin{figure}[h]
    \centering
    \includegraphics[width=1\linewidth]{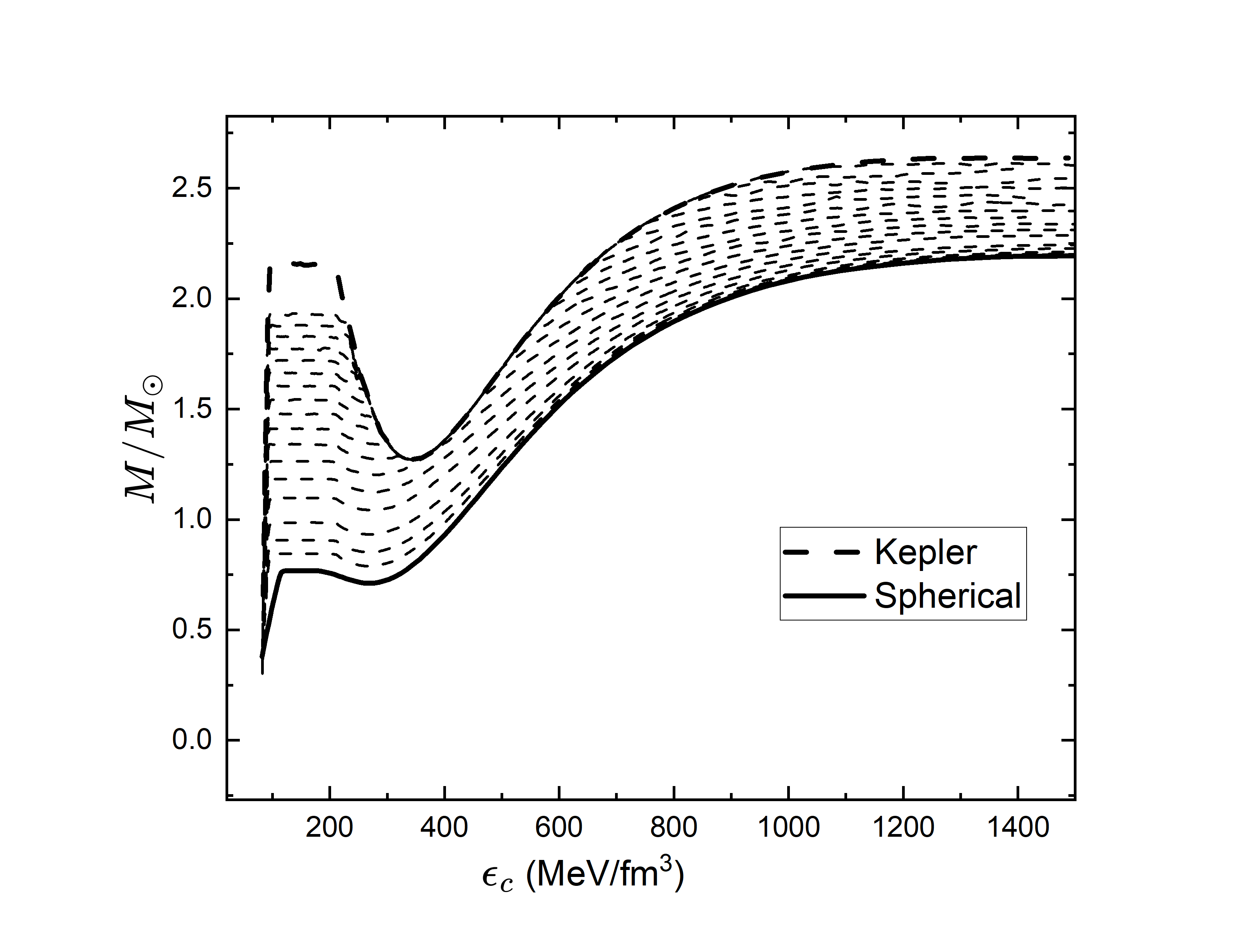}
    \caption{Same as Fig.~\ref{fig:JseqEOSA} except for EOS C. } \label{fig:JseqEOSC}
\end{figure}

We now discuss a universal relation for the supramassive configurations, which are stars more massive than their non-rotating counterparts. In order to do that, we must analyze several sequences of constant angular momentum ($J$) for the three EOS's studied. Such sequences are shown in Figs.~\ref{fig:JseqEOSA}, \ref{fig:JseqEOSB} and \ref{fig:JseqEOSC}.

We now turn our attention to the maximum mass configuration for the sequences depicted in Figs.~\ref{fig:JseqEOSA}--\ref{fig:JseqEOSC}, which we define as $M_\uparrow$. For the spherically symmetric sequence, this is nothing less than the maximum TOV mass ($M_{max}^{TOV}$), whereas for the Kepler sequence this star represents the supramassive limit, i.e. the upper limit mass that a (rigidly) rotating star may have. The authors of \cite{Breu2016,Bozzola2019} have found, for hadronic stars, a universal relation between $M_\uparrow$ normalized by $M_{max}^{TOV}$ and the corresponding star's angular momentum $J$, normalized by the maximum possible angular momentum ($J_{Max}$), which is written as
\begin{equation}
    \frac{M_\uparrow}{M_{Max}^{TOV}} = 1 + 0.1316 \left( \frac{J}{J_{Max}}\right)^2 + 0.0711 \left( \frac{J}{J_{Max}}\right)^4
\label{eq14}
\end{equation}

We perform a similar study, now considering the models under investigation in this work. In order to do that, we have numerically found the maximum mass configuration for the different constant angular momentum sequences (normalized by $J_{Max}$) , normalizing it by their corresponding $M_{max}^{TOV}$. With this quantity, we have found that the best fit curve for inverted hybrid stars is given by
\begin{equation}
    \frac{M_\uparrow}{M_{Max}^{TOV}} = 1 + a\left( \frac{J}{J_{Max}}\right)^2 - b \left( \frac{J}{J_{Max}}\right)^4,
\end{equation}
where the constants assume the values $a = 0.3186 \pm0.017$ and $b=0.0886\pm0.024$. 

\begin{figure}[h]
    \centering
    \includegraphics[width=0.95\linewidth]{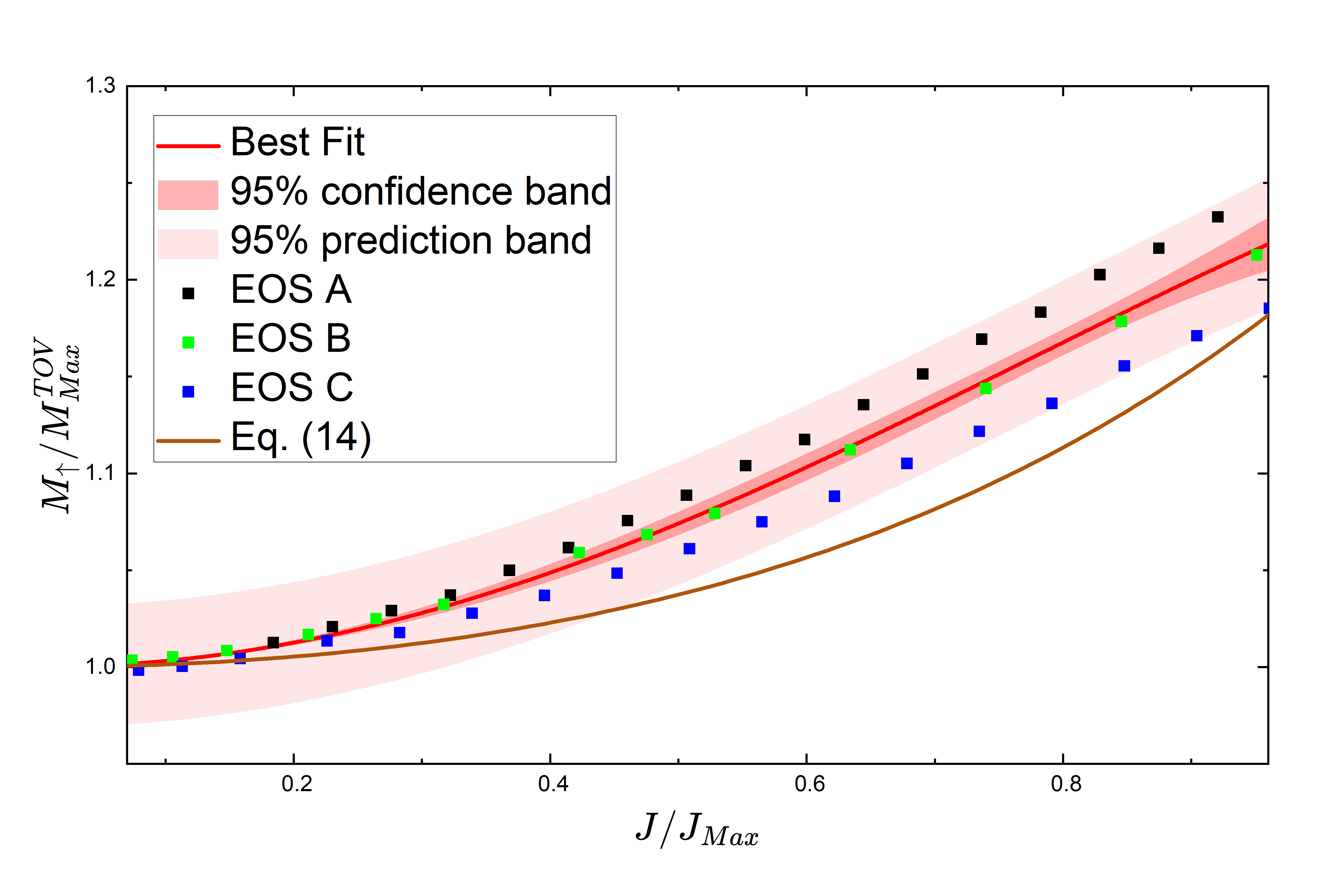}
    \caption{Normalized gravitational mass of the top mass as a function of the corresponding normalized stellar angular momentum, for the three different EOSs studied. Also shown are the best-fit curve and the 95\% confidence and prediction bands. For comparison purposes, we also show the curve for Eq.~(\ref{eq14}). } \label{fig:UniFIT}
\end{figure}

Our results are shown graphically in Fig.~\ref{fig:UniFIT}, where besides the numerical data, we also show the best-fit curve as well as the confidence and prediction band of $95\%$.

The results depicted in Fig.~\ref{fig:UniFIT} indicate that most likely IHSs also satisfy universal relations, at the very least for the supramassive limit, although this would need to be confirmed by a more exhaustive study with a larger set of EOS for IHSs and also investigating other universal relations.

Now we must also spend some time discussing the relationship between the star's moment of inertia ($\hat{I}$) and the spin-induced quadrupole moment ($\hat{Q}$). The authors of \cite{yagi2013lovea,yagi2013loveb} have shown that, for slowly rotating neutron stars, there is a universal relation connecting $\hat{I} = I/M^3$ and $\hat{Q} = -Q/M^3a^2$ (where $a = J/M^2$) given by
\begin{equation}
\begin{split}
    \ln{\hat{I}_{YY}} = 1.35 + 0.697\ln{\hat{Q}} - 0.143 \ln{\hat{Q}^2} + \\ 9.94\times10^{-2} \ln{\hat{Q}^3} - 1.24\times10^{-2} \ln{\hat{Q}^4}. \label{IQfit}
    \end{split}
\end{equation}

\begin{figure}[h]
    \centering
    \includegraphics[width=0.95\linewidth]{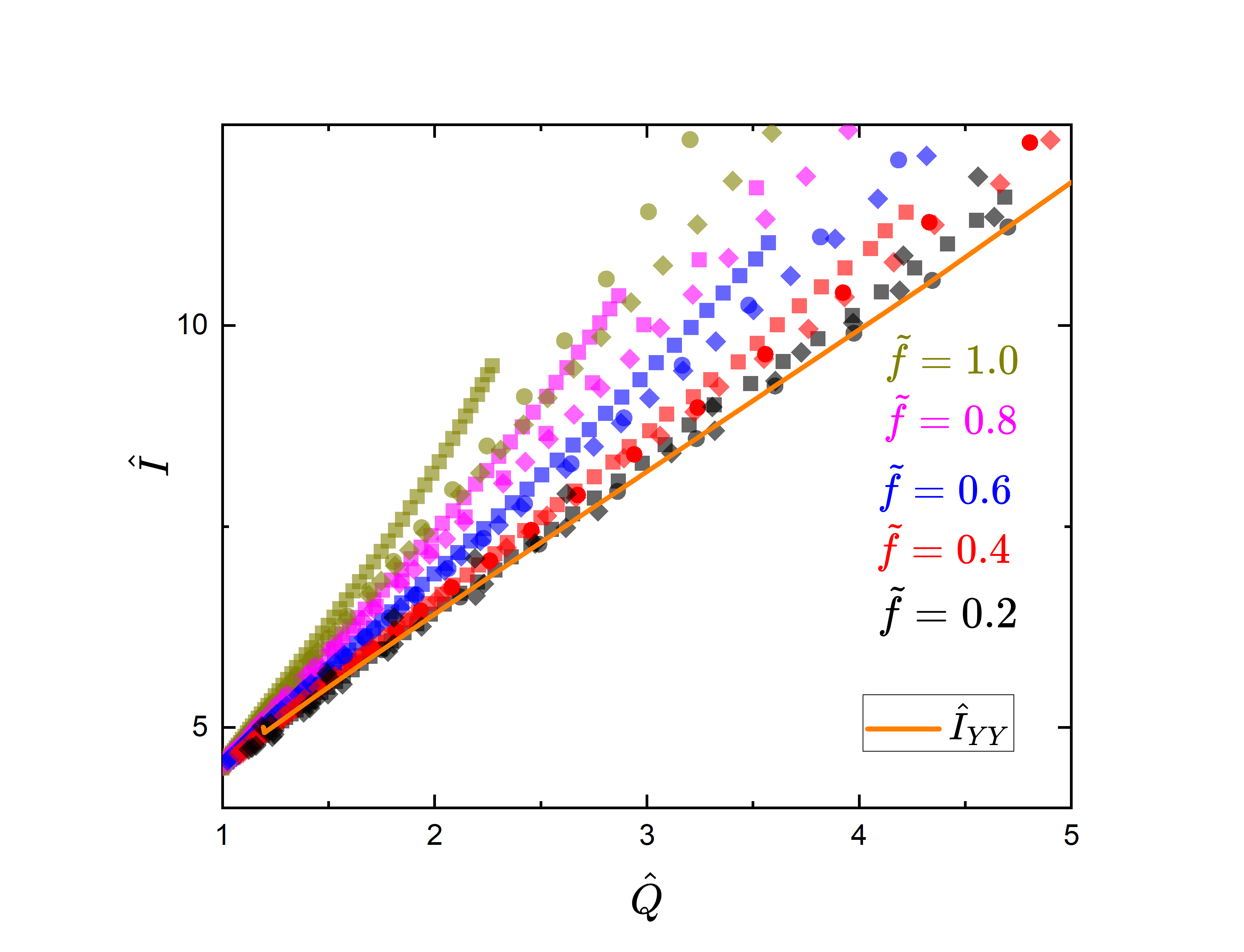}
    \caption{$\hat{I}$ as a function of $\hat{Q}$ for stars with different values of the parameter $\tilde{f}$ as indicated by the figure. Different symbols denote different EOS, with squares representing EOS A, circles EOS B, and diamonds EOS C. Also shown is the relation found by \cite{yagi2013lovea,yagi2013loveb} for slowly rotating neutron stars } \label{fig:IhatQhat}
\end{figure}

The work presented in \cite{doneva2013breakdown} has extended this UR study to rapidly rotating neutron stars, and found a breakdown of universality in the rapidly rotating case. The authors of \cite{chakrabarti2014q} made further refinements showing that a much better universality is achieved 
if certain dimensionless parameters characterizing the magnitude of rotation are used, such as $\tilde{f}= 20 R\Omega$, where $R$ is the equatorial radius. For this part of our study we follow the approach of \cite{chakrabarti2014q} and calculate the relationship between $\hat{I}$ and $\hat{Q}$ for constant sequences $\tilde{f}$. This result is shown in Fig.~\ref{fig:IhatQhat}, where we also compare the $\hat{I}-\hat{Q}$ relation with the fit obtained by \cite{yagi2013lovea,yagi2013loveb}.

We can surmise by the results of Fig.~\ref{fig:IhatQhat} that IHSs behave similarly to the other models studied in reference \cite{doneva2013breakdown}, where it was shown that rapidly rotating stars deviate from the fit found in \cite{yagi2013lovea,yagi2013loveb}, especially for higher values of $\hat{Q}$. 

\begin{figure}[h]
    \centering
    \includegraphics[width=0.95\linewidth]{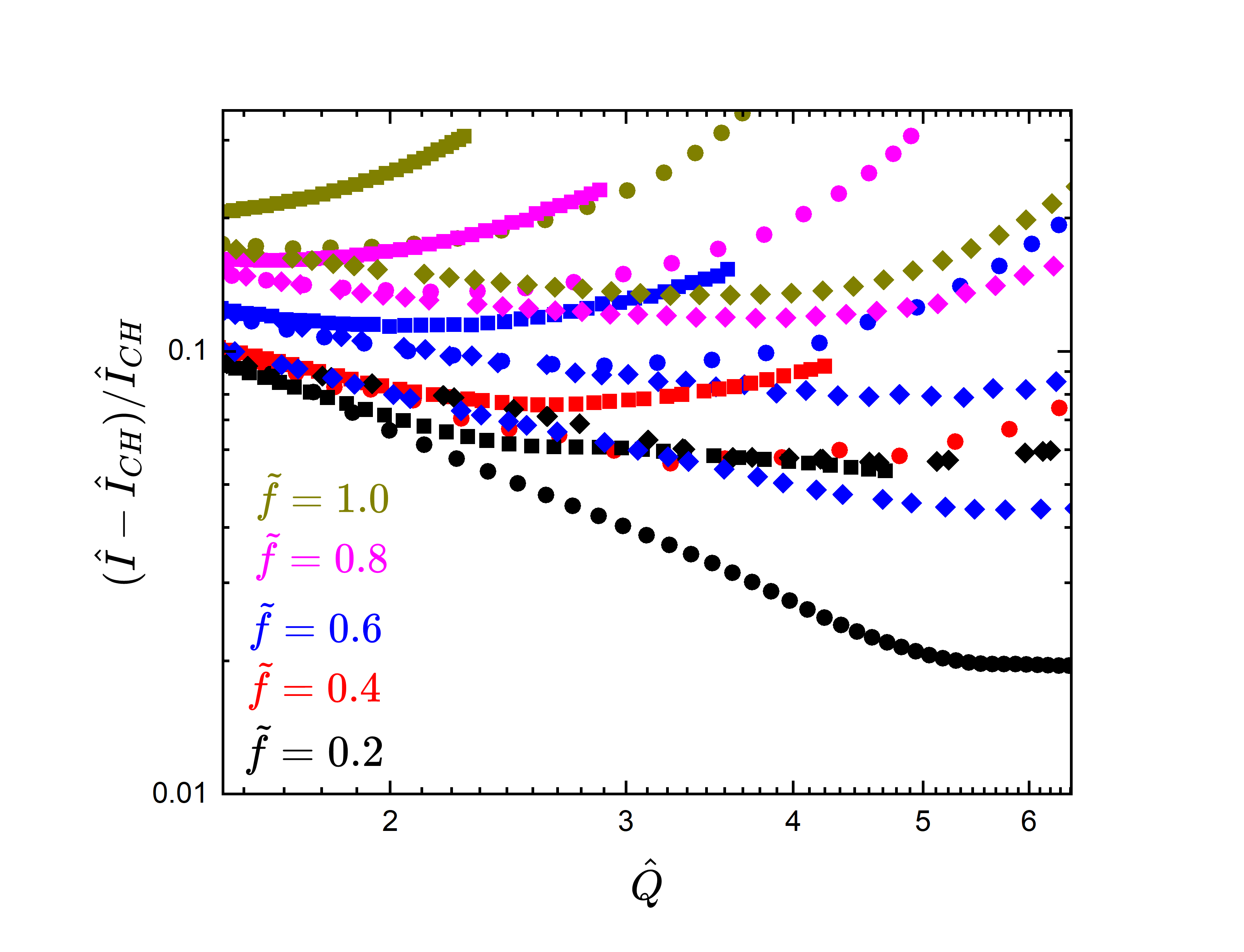}
    \caption{ Relative deviation of the calculated $\hat{I}$ to the fit $\hat{I}_{CH}$ for stars with different values of the parameter $\tilde{f}$ as indicated by the figure. Different symbols denote different EOS, with squares representing EOS A, circles EOS B, and diamonds EOS C.  } \label{fig:Irel}
\end{figure}

Ref.~\cite{chakrabarti2014q} extended the $\hat{I}-\hat{Q}$ fitting relation eq. (\ref{IQfit}) with an expansion of the parameter $\tilde{f}$ to achieve better universality in the rapidly rotating case. Therefore, we now compare the numerical $\hat{I}-\hat{Q}$ found for IHSs (Fig.~\ref{fig:IhatQhat}) with the fit proposed in \cite{chakrabarti2014q} (labeled here $I_{CH}$), by showing the relative deviation between these quantities in Fig.~\ref{fig:Irel}.

The results shown in Fig.~\ref{fig:Irel} suggest that IHSs adhere to the universal behavior outlined in \cite{chakrabarti2014q}, particularly for smaller values of $\hat{Q}$ and stars with low $\tilde{f}$. However, there are relatively larger discrepancies for higher values of $\hat{Q}$. These findings diverge from those reported by \cite{paschalidis2018implications}, where the $\hat{I}-\hat{Q}$ relation for different (ordinary) hybrid stars revealed a smaller relative discrepancy.

We speculate that this variation arises from the absence of a nuclear crust in IHSs, which have an outer region composed of quark matter, unlike ordinary neutron and hybrid stars. This lack of a crust may explain the observed discrepancies in high $\hat{Q}$ stars, which are associated with smaller central densities where low-density quark matter plays a more significant role.

To shed further light on this behavior, we plan to investigate the properties of the $\hat{I}-\hat{Q}$ relation for bare quark stars compared to neutron stars and IHSs in future studies.

\section{Conclusions}
In this research, we built upon the work of \cite{Zhang2023,Zhang2024,Holdom2018} which has put forth the intriguing possibility that quark matter can be more stable than hadronic matter at low pressure and be less stable at higher pressure. This provocative possibility gives rise to inverted hybrid stars or cross stars by other names, in which their outer layer is composed of quark matter while the higher-density core is made of hadrons, wherein the quarks are confined. 

The studies presented in \cite{Zhang2023, Zhang2024} comprehensively examine the characteristics of cross stars, such as mass, radius, tidal deformability, and oscillations. Our research aims to extend their findings by investigating the attributes of rotating IHSs, including the potential evolutionary trajectories of objects with constant baryonic mass. We find that such studies are garnering renewed interest, especially considering the possibility of post-merger objects with high angular momentum. We have then analyzed three benchmark models for IHSs, each representing a different ``depth" of the $ud$QM to hadronic matter phase transition: deep (EOS A), intermediate (EOS B), and shallow (EOS C). All equations of state presuppose the APR EOS for the hadronic matter. We have computed various families of rotating neutron stars across all models to gain a comprehensive understanding of their properties.

Our research reveals that IHSs rotating at their Kepler frequencies typically exhibit a higher mass and a larger circumferential radius, as anticipated. We have also observed an interesting behavior regarding twin configurations, where twin stars are compact objects with identical mass but differing radii, a characteristic often observed in hybrid stars.  Our results indicate that in configurations with high angular momentum, such as those in the Kepler sequence, the likelihood of potential twin configurations increases significantly. Moreover, EOS B and EOS C start to exhibit twin stars within their Kepler sequences, despite not presenting such configurations in the non-rotating case.

We have also devoted considerable effort to studying sequences of constant baryonic mass. These sequences are characterized by different stellar configurations that share the same baryonic mass but exhibit varying angular momentum. These configurations reach their extremes at the Kepler sequence, beyond which no hydrostatically stable configurations exist, and at the spherical sequence, which represents the minimum possible angular momentum. Sequences of constant baryonic mass between these two extremes represent potential paths of rotational evolution that stars may follow as they transition from high angular momentum to spherical, non-rotating configurations. 
We have shown that although we find a sequence of constant baryon mass connecting stars at the Kepler frequency to their non-rotating counterparts, not all stars in these sequences are viable. While all of them satisfy the hydrostatic equilibrium conditions, some of them are not stable against oscillations. These unstable configurations occur due to the onset of phase transition during the spin-down, which happens as the central density increases during the spin-down, triggering the phase transition within the star. This leads to the possibility of a mini-collapse, occurring in the spin-down of stars with high angular momentum, such as those studied in Ref. \cite{haensel2016rotating}. We have estimated that for the current model, these mini-collapses are associated with the release of $\sim 10^{51}$ erg of energy, part of which would be released in the form of electromagnetic and/or gravitational radiation. 

We have further investigated the properties of constant baryonic mass during spin-down sequences, particularly the phenomena of ``back-bending". This phenomenon is characterized by an increase in rotational frequency that coincides with a reduction in angular momentum for stars nearing the onset of phase transition. Typically observed in stars undergoing phase transition during spin-down, this phenomenon arises from the need for a rapid increase in rotational frequency to compensate for a substantial decrease in the moment of inertia during the phase transition. This is usually manifested as an ‘S’ shape in the angular momentum graph. However, due to the unique nature of IHSs, which have a quark phase in the outer layers, the back-bending for these stars is exhibited in a rather distinctive way, behaving more akin to what can be described as a ‘back-kink’ due to its sharp shape. Although admittedly remote given our limited observational data on the angular momentum and moment of inertia of neutron stars, there may be a possibility of distinguishing between IHSs and ordinary hybrid stars if more observational data on the relevant quantities become available in the future.

Furthermore, we have calculated the braking index of IHSs during spin-down. Our results indicate that for low-frequency stars, when their shape is nearly spherical, the braking index approximates the canonical value of three, as expected. For higher frequencies, where the star’s shape and moment of inertia deviate significantly from those of spherical stars, the effects of back bending become apparent on the braking index. This index substantially increases (in absolute value) in the back-bending region. These observations could potentially serve as evidence for probing the star's internal composition.

We have also explored whether universal relations for the supramassive mass can be found across all studied models. This relation is significant as they relate the maximum mass of a rigidly rotating star to the TOV mass limit and the star's angular momentum. We have identified such a connection with a relatively high level of numerical confidence. We believe that by investigating more IHSs EOS models, we can achieve an even more precise correlation. Our future plans include continuing this investigation with more microscopic models for IHSs and expanding our search for universal relations for IHSs.

Furthermore, concerning universal relations, our studies on the $\hat{I}-\hat{Q}$ relation of IHSs have demonstrated that they exhibit similar behaviors to the universal relations found for rapidly rotating stars, albeit only for low values of $\hat{Q}$. Beyond these values, discrepancies emerge. We attribute this behavior to the quark matter crust of IHSs and plan to investigate this phenomenon further in future research.

With this work, our objective has been to build upon previous studies of inverted hybrid stars by incorporating the significant aspect of rotation. Our research has unveiled intriguing phenomena that occur in this new type of compact star when angular momentum is taken into account, some of which could potentially serve as observational markers with the emergence of new data. We believe that our work enriches the studies presented by \cite{Zhang2023,Holdom2018}, introducing an additional and captivating layer of complexity. We intend to delve deeper into the fascinating concept of IHSs by investigating differential rotation and magnetic fields, a study that is currently in progress and will be detailed in our future works.\\

\begin{acknowledgments}
R.N. acknowledges financial support by FAPERJ under project Proc. No. E-26/201.432/2021. R.N. extends his gratitude to Prof. Renxin Xu for the warm hospitality and pleasant atmosphere provided during the visit to PKU in October 2023. C.~Z. is supported by the Jockey Club Institute for Advanced Study at The Hong Kong University of Science and Technology. 
\end{acknowledgments}

\nocite{*}

\bibliography{apssamp}

\end{document}